\documentclass[11pt,a4paper]{article}
\usepackage{tikz}
\usepackage[compat=1.0.0]{tikz-feynman}
\usepackage{authblk}
\usepackage{placeins}
\usepackage{epsfig}
\usepackage{subfigure}
\usepackage{cancel}
\usepackage{comment}
\usepackage{comment}
\usepackage{amsmath}
\usepackage{siunitx}
\usepackage{amssymb}
\usepackage{mathtools}
\usepackage{graphicx}
\usepackage{subcaption}  
\usepackage{caption}      
\usepackage{float}        
\usepackage{MnSymbol}
\graphicspath{{plots/}}
\usepackage{doi}           % makes DOI fields work
\usepackage{hyperref}
\usepackage{csquotes}
\usepackage{multirow}
\usepackage[left=.45in,right=.45in,top=.6in,bottom=.6in,headheight=14.5pt]{geometry}
\usepackage{multirow}
\usepackage{colortbl}
\usepackage{xcolor}
\usepackage{pdflscape}
\usepackage{orcidlink}
\usepackage[toc,page]{appendix}
\usepackage[left=.45in,right=.45in,top=.6in,bottom=.6in,headheight=14.5pt]{geometry}
\usepackage{fmtcount} % for \ordinalnum
\usepackage{array,multirow}
\newcolumntype{C}[1]{>{\centering\arraybackslash}m{#1}}
\usepackage{epsfig}
\usepackage{float}
\usepackage{geometry}
\usepackage{amsmath}
\usepackage{amssymb,tikz-feynman}
\usepackage{cite}
\usepackage{ctable}
\newgeometry{vmargin={10mm,20mm},hmargin={19mm,19mm}}

\title{\textbf{Tri-Resonant Leptogenesis  in a Non-Holomorphic Modular A$_4$ Scotogenic Model}}
\author{Tapender\thanks{tapenderphy@gmail.com}\;\;\orcidlink{0009-0006-4109-528X} and Surender Verma\thanks{s\_7verma@hpcu.ac.in}\;\;\orcidlink{0000-0002-5671-5369}}
\date{}
\affil{Department of Physics and Astronomical Science, Central University of Himachal Pradesh, Dharamshala-176215, INDIA}

\begin{document}	
\maketitle

\begin{abstract}

\noindent We investigate low-scale baryogenesis \textit{via} tri-resonant leptogenesis within the scotogenic model with a scalar dark matter  embedded in a non-holomorphic modular $A_4$ symmetry framework. We demonstrate that the model naturally accommodates three nearly degenerate right-handed (RH) neutrinos when they are assigned to the triplet representation of $A_4$. The near degeneracy originates from treating the symmetric contribution to the Majorana mass matrix, arising from the $\mathbf{3}\otimes\mathbf{3}$ decomposition of $A_4$, as a small perturbation to the dominant singlet contribution. Generalized CP (gCP) symmetry is imposed in the model, rendering the complex modulus $\tau$ as the sole source of CP violation which is highly constrained by neutrino oscillation data, making the framework highly predictive. 
% The model yields definite light neutrino mass hierarchy dependet predictions for the atmospheric mixing angle $\theta_{23}$, CP-violating phases, the effective Majorana mass $m_{ee}$, and the sum of neutrino masses $\sum_i m_i$, which depend on the hierarchy of light neutrino masses. 
% This model has various predictions which depends on the mass hierarchy of light neutrino masses.
In particular, for the inverted hierarchy (IH), the predicted $3\sigma$ range of $\theta_{23}$ lies in the lower octant close to maximal value while the Dirac CP phase $\delta_{\mathrm{CP}}$ and the Majorana phase $\alpha_{21}$ are predicted to lie close to $0^\circ$ or $360^\circ$. Also, in this case, predicted values of $m_{ee}$ and $\sum_i m_i$ can be tested and constrained by future neutrinoless double beta decay $(0\nu\beta\beta)$ experiments, as well as by cosmological observations, particularly DESI+BAO and Planck data. In fact DESI+BAO disallows IH in the model. We further show that successful baryogenesis can be achieved for both normal hierarchy (NH) and inverted hierarchy (IH) of light neutrino masses with RH neutrino masses as low as $537~\mathrm{GeV}$ rendering this scenario experimentally testable.  For NH, RH neutrino mass degeneracy of $\mathcal{O}(10^{-7}\!-\!10^{-6})$ is required, while for IH a stronger degeneracy of $\mathcal{O}(10^{-8})$ is needed. Remarkably, in the NH case, successful baryogenesis can occur even in the deep washout regime with decay parameters of $\mathcal{O}(10^{5})$ owing to the tri-resonant enhancement of the CP asymmetry and the inclusion of flavor effects.
% Overall, this framework provides a highly constrained and testable scenario with multiple predictions accessible to future neutrino, cosmological, and collider experiments.
\end{abstract}

\section{Introduction}
The Standard Model (SM) of particle physics is one of the most successful theories developed to describe the fundamental constituents of nature and their interactions. Despite its remarkable achievements, it is now well established that the SM is incomplete. In particular, it treats neutrinos as massless particles, whereas compelling experimental evidence from neutrino oscillation experiments confirms that neutrinos are massive \cite{Ahmad:2002jz,An:2012eh,Abe:2011fz,SNO:2001kpb,KamLAND:2002uet,Super-Kamiokande:1998kpq}. Furthermore, the SM fails to account for the existence of dark matter in the Universe and cannot explain the observed matter--antimatter asymmetry.
To address these shortcomings, the SM is often extended by introducing additional fields and symmetries. One well-motivated extension is the scotogenic model proposed by Ma~\cite{Ma:2006km}, which provides a unified framework for explaining both the origin of neutrino masses and the nature of dark matter~\cite{Racker:2014yfa,Racker:2024fpn,Hugle:2018qbw,Borah:2018rca,Sarma:2020msa,Mahanta:2019gfe,Mahanta:2019sfo,Racker:2013lua,Kashiwase:2013uy,Kashiwase:2012xd,Suematsu:2013owa,Thapa:2022fhv,Sarma:2021acj,Boruah:2021qlf,Tapender:2024ktc,Tapender:2025xoj,Singh:2023eye,Borah:2020ivi,Arora:2024lpq,Tapender:2026jji,Behera:2020lpd,Singirala:2023zos,Avnish:2023zbl,Ahriche:2022bpx,Borah:2023hqw,Arora:2022hza,Ankush:2021opd}. However, generating the observed baryon asymmetry of the Universe through leptogenesis \cite{Fukugita:1986hr,Davidson:2008bu,Buchmuller:2005eh,Fong:2012buy,Pilaftsis:2009pk,Buchmuller:2004nz} within the scotogenic model presents significant challenges, particularly when the relevant mass scale lies at or below the TeV scale. These challenges mainly arise from strong washout effects induced by the large Yukawa couplings required to satisfy neutrino oscillation data. Various mechanisms have been proposed in the literature to circumvent this issue and successfully realize the required baryon asymmetry without extended the scotogenic model~\cite{Racker:2014yfa,Racker:2024fpn,Hugle:2018qbw,Borah:2018rca,Sarma:2020msa,Mahanta:2019gfe,Mahanta:2019sfo,Racker:2013lua,Kashiwase:2013uy,Kashiwase:2012xd,Suematsu:2013owa,Thapa:2022fhv,Sarma:2021acj,Boruah:2021qlf} and with extension~\cite{Singh:2023eye,Borah:2020ivi,Arora:2024lpq}.
One possible approach is to reduce the Yukawa couplings in order to suppress the washout effects. However, this reduction simultaneously diminishes the generated CP asymmetry, thereby resulting in an insufficient lepton asymmetry.
This limitation can be overcome by invoking resonant leptogenesis~\cite{Pilaftsis:1997jf,Pilaftsis:2003gt,Pilaftsis:2005rv}, where a small mass splitting among the decaying particles enhances the CP asymmetry. Consequently, it becomes possible to identify regions of parameter space where neutrino oscillation data, dark matter relic density constraints, and the observed baryon asymmetry of the Universe can be simultaneously satisfied.
Previous studies employing resonant leptogenesis within the scotogenic framework have successfully achieved this goal, but typically at mass scales above 1~TeV~\cite{Kashiwase:2013uy,Kashiwase:2012xd,Suematsu:2013owa,Thapa:2022fhv}. In the present work, we aim to lower the mass scale below 1~TeV without resorting to extremely small Yukawa couplings. As a result, the model operates in the strong washout regime, with the decay parameter reaching values as large as $10^{5}$. This scenario is realized within the framework of tri-resonant leptogenesis~\cite{Pilaftsis:2003gt,daSilva:2022mrx,Pilaftsis:2023snk,Liu:2024eki}, which requires three nearly degenerate right-handed (RH) neutrinos. Achieving such a spectrum poses a significant challenge from a model-building perspective. 
% We emphasize, however, that resonant leptogenesis is not the only viable approach to low-scale baryogenesis in the scotogenic model; alternative mechanisms exist that do not rely on resonant enhancement~\cite{}.

In this work, we implement the scotogenic model within the framework of recently proposed non-holomorphic modular $A_4$ symmetry~\cite{Qu:2024rns}. In this setup, it is not necessary to impose an additional $Z_2$ symmetry, as appropriate modular weight assignments forbid unwanted interactions. Unlike the conventional holomorphic modular symmetry~\cite{Feruglio:2017spp}, non-holomorphic modular symmetry can be applied to non-supersymmetric (non-SUSY) frameworks. Moreover, it allows for modular forms with negative modular weights, thereby increasing the model-building flexibility while preserving the stringent constraints imposed by modular invariance.
In these constructions, Yukawa couplings are not arbitrary parameters but are determined by modular forms, which are functions of the complex modulus $\tau$. If a generalized CP symmetry~\cite{PhysRev.148.1385,Bree:2024edl,Tapender:2023kdk,Tapender:2025cfc,Priya:2025khf} is imposed consistently with the non-holomorphic modular symmetry~\cite{Novichkov:2019sqv,Qu:2024rns}, the modulus $\tau$ becomes the sole complex parameter of the model and thus the only source of CP violation. Since CP violation plays a crucial role in leptogenesis~\cite{Andrei:1991}, this feature strongly constrains the CP-violating phases and renders the framework highly predictive.
There exist previous studies of leptogenesis in holomorphic~\cite{Asaka:2019vev,Ding:2022bzs,Behera:2022wco,Pathak:2025zdp,Kashav:2022kpk,Marciano:2024nwm,Behera:2020sfe,Mohanta:2023tzf,Abhishek:2025ety,Kang:2022psa,Mishra:2022egy,Singh:2024imk,Pathak:2025fpo,Gogoi:2023jzl,Kashav:2021zir} and non-holomorphic~\cite{Nanda:2025lem,Kumar:2025bfe,Priya:2025wdm,Nasri:2026nbf} modular frameworks. As we focus on low-scale leptogenesis, flavor effects become important. Accordingly, we solve the flavor-dependent density matrix equations, which properly account for flavor dynamics in the evolution of the RH neutrino and lepton number densities~\cite{Abada:2006fw,Barbieri:1999ma,DeSimone:2006nrs,Blanchet:2006ch,Blanchet:2011xq,Moffat:2018wke,Granelli:2021fyc}.
Within this framework, we demonstrate that three nearly degenerate RH neutrinos can naturally arise when they are assigned to the triplet representation of $A_4$. The near degeneracy originates from treating the symmetric contribution to the Majorana mass matrix, arising from the $\mathbf{3}\otimes\mathbf{3}$ decomposition of $A_4$, as a small perturbation to the dominant singlet contribution. The singlet term exhibits a $(2,3)$ symmetry structure, which is broken by the symmetric triplet contribution when treated perturbatively.
We note that a scotogenic model based on non-holomorphic modular $A_4$ symmetry has also been studied in Ref.~\cite{Nomura:2024vzw}. However, leptogenesis was not considered in that work, and the modular representations differ from those adopted here. Consequently, our framework leads to distinct phenomenological predictions.

This article is organized as follows. In Section~\ref{sec2}, we describe the model and its theoretical framework. In Section~\ref{sec3}, we analytically demonstrate how three nearly degenerate RH neutrinos naturally emerge in this setup. In Section~\ref{sec4}, we present the numerical analysis and discuss neutrino phenomenology. Section~\ref{sec5} is devoted to the formalism of tri-resonant leptogenesis, where we solve the flavor-dependent density matrix equations to study the evolution of RH neutrino and lepton number densities. Finally, in Section~\ref{sec6}, we summarize our results and present our conclusions.

\section{Model and Framework}\label{sec2}

We implement the scotogenic model within a non-holomorphic modular \(A_4\) symmetry framework. 
The particle content of the model is identical to that of Ma’s original scotogenic model~\cite{Ma:2006km}. 
In addition to the Standard Model field content namely, the left-handed (LH) lepton doublets $L_L$, the right-handed charged lepton singlets $(e_R, \mu_R, \tau_R)$ and the Higgs doublet $H$ the model contains three RH neutrinos $N_{\alpha}$ ($\alpha = 1, 2, 3$) and an inert scalar doublet $\eta$.

Instead of imposing a discrete \(Z_2\) symmetry to forbid the tree-level neutrino mass terms, 
modular weights are employed for this purpose. Fields with odd modular weights effectively behave 
as \(Z_2\)-odd particles and therefore belong to the dark sector of the model. 
The assignments of all particles under the Standard Model gauge group and the non-holomorphic 
modular \(A_4\) symmetry along with their corresponding modular weights are summarized in 
Table~\ref{tab1}. In Table~\ref{tab2}, we list the modular forms used in the model together with 
their \(A_4\) representations and modular weights.

% We have implemented the scotogenic model in the  non-holomorphic A$_4$ modular symmetry framework. The particle content of the model  is same as in the Ma's scotogenic model \cite{}. Apart from the standard model field   content we have three right-handed neutrinos (N$_{R_i}$, $i=1,\;2,\;3$) and one inert scalar doublet $\eta$. Instead of using the discrete $Z_2$ symmetry to forbid the tree-level mass term for the neutrinos modular weights are used. All particles with odd modular weight are like Z$_2$ odd particles and hence belongs to dark sector of the model. The various assignments of the particles under SM and non-holomorphic modular A$_4$ group along with their weights are shown in table \ref{tab1}. In Table \ref{tab2}, we  have shown the modular forms used in this model along with their A$_4$ representation and modular weights.    

% \begin{table}[h]
% 	\centering
% 	\begin{tabular}{cccccc}
% 		Fields   &$L_{e,\mu,\tau}$&$e_R,\mu_R,\tau_R$ & $N_R$&$\eta$& H \\ \hline
% 		SU(2)$_L$   &2  &  1& 2 & 2& 2\\
% 		U(1)$_Y$ & $-1$& $-2$& $-1$& 1& 1\\
% 		% U(1)$_{B-L}$& -1 & -1 & -1 & -1 & -1 & 0 &0 \\
% 		A$_4$ & $1,\; 1^{\prime },\;1^{\prime \prime}$&$1,\; 1^{\prime \prime},\; 1^\prime$& 3 & 1 &1 \\
% 		$-k_I$& $2$& $-2$& $-1$& $1$& 0\\
% 	\end{tabular}
% 	\caption{Charge assignments for various fields under different symmetry groups.}
% 	\label{tab1}
% \end{table}

\begin{table}[t]
	\centering
	\begin{tabular}{|c|c|c|c|c|c|}
    \hline
		Fields   &$L_{L}$&$e_R,\mu_R,\tau_R$ & $N$&$\eta$& H \\ \hline
		SU(2)$_L$   &2  &  1& 1 & 2& 2\\
		U(1)$_Y$ & $-1$& $-2$& $-1$ & $-1$ & 1\\
		% U(1)$_{B-L}$& -1 & -1 & -1 & -1 & -1 & 0 &0 \\
		A$_4$ & $3$&$1,\; 1^{\prime },\; 1^{\prime \prime}$& 3 & 1 &1 \\
		$-k_I$& $-2$& $2$& $-1$& $1$& 0\\
        \hline
	\end{tabular}
	\caption{Charge assignments for various fields under different symmetry groups.}
	\label{tab1}
\end{table}

\begin{table}[t]
	\centering
	\begin{tabular}{|c|c|c|c| }
		\hline
		& $ Y_{1}^{(2)}$& $Y_3^{(0)}$&$Y_3^{(2)}$ \\ 
		\hline
		$A_4$ & 1 & 3&$3$ \\
		$k_Y$ & 2 & 0&$2$ \\
        \hline
	\end{tabular}
	\caption{Modular forms, their A$_4$ representations and modular weights.}
	\label{tab2}
\end{table}

% \begin{table}[h]
% 	\centering
% 	\begin{tabular}{c c c }
% 		\hline
% 		& $Y_3^{(-2)}$ & $Y_3^{(2)}$ \\ 
% 		\hline
% 		$A_4$ & $3$ & $3$ \\
% 		$-k_I$ & $-2$ & $2$ \\
% 	\end{tabular}
% 	\caption{Modular forms, their A$_4$ representations and modular weights.}
% 	\label{tab2}
% \end{table}

The Yukawa Lagrangian for the model is\footnote{Here, the notation $[\,\dots\,]_r$ indicates that the object inside the brackets transforms as the representation $r$ of the $A_4$ group.}
\begin{equation}
	\begin{aligned}
		\mathcal{L}_Y =\;&
		% ---- Charged lepton Yukawas ----
			\beta_e [Y_3^{(0)}\overline{L}_L]_1 H e_R 
			+\beta_\mu [Y_3^{(0)}\overline{L}_L]_{1^{\prime \prime} }H \mu_R 
			+ \beta_\tau [Y_3^{(0)}\overline{L}_L]_{1^\prime} H \tau_R
		 \\[6pt]
		&+\;
			 Y_{1}^{(2)} \beta_1 [\overline{L}_L N]_1 \eta + Y_{3}^{(2)}\!\left( \beta_2
			[\overline{L}_L N]_{3_S} \eta 
			+ \beta_3 [\overline{L}_L N]_{3_A} \eta \right)
		 \\[6pt]
		&+\;  Y_{1}^{(2)}\, k_1 [\overline{N^C} N]_{1} +
			Y_{3}^{(2)}\, k_2 [\overline{N^C} N]_{3_S}
		\;+\; \text{h.c.}.
	\end{aligned}
\end{equation}
where $\beta_l$ ($l = e, \mu, \tau$), $\beta_i$ ($i = 1, 2, 3$), and $k_i$ ($i = 1, 2$) are coupling constants. We consider a generalized CP (GCP) symmetry~\cite{Novichkov:2019sqv,Qu:2024rns}, under which all coupling constants are real and the modulus $\tau$ is the only complex parameter responsible for CP violation. The $A_4$ triplet modular forms can be written as $Y_3^{(0)}=(Y^0_1,Y^0_2,Y^0_3)$ and $Y_3^{(2)}=(Y^2_1,Y^2_2,Y^2_3)$.

After applying the rules of  $A_4$ tensor product~\cite{Ishimori:2010au}, the relevant Yukawa Lagrangian can be written as
\begin{equation}
	\mathcal{L}_Y = 
	Y_c\, \overline{L}_L H l_R  
	+ Y_{\eta} \, \overline{L}_L \eta N 
	+ M_{N} \, \overline{N^C} N 
	+ \text{h.c.},
	\label{eq:LY_int}
\end{equation}
where $L_L$ and $l_R$ denote the LH and RH $SU(2)_L$ lepton doublets and singlets, respectively. The symbol $Y_c$ represents the charged-lepton Yukawa coupling matrix, $Y_{\eta}$ denotes the Yukawa coupling matrix involving the RH neutrinos, the inert scalar doublet and the LH lepton doublets and $M_{N}$ is the bare mass matrix of the RH neutrinos.

The charged-lepton Yukawa coupling matrix is non-diagonal and has the following form
\begin{equation}
	Y_c =
\begin{pmatrix}
		\beta_e Y_{1}^0 & \beta_\mu Y_{3}^0 & \beta_\tau  Y_{2}^0 \\
		\beta_e Y_{3}^0 & \beta_\mu Y_{2}^0 & \beta_\tau Y_{1}^0 \\
		\beta_e Y_{2}^0 & \beta_\mu Y_{1}^0 & \beta_\tau Y_{3}^0
	\end{pmatrix},
	\label{eq:Yc}
\end{equation}
and the Yukawa coupling matrix $Y_\eta$  is of the form
	\begin{equation}
					Y_\eta =
					\begin{pmatrix}
						\beta_1  Y_{1}^{(2)} + 2 Y_{1}^2 \beta_2 & Y_{3}^2 (-\beta_2 + \beta_3) & Y_{2}^2 (-\beta_2 - \beta_3) \\
						Y_{3}^2 (-\beta_2 - \beta_3) & 2 Y_{2}^2 \beta_2 & \beta_1  Y_{1}^{(2)}  + Y_{1}^2 (-\beta_2 + \beta_3) \\
						Y_{2}^2 (-\beta_2 + \beta_3) & \beta_1Y_{1}^{(2)} + Y_{1}^2 (-\beta_2 - \beta_3) & 2 Y_{3}^2 \beta_2
					\end{pmatrix}.
				\end{equation}

The bare mass matrix for the right-handed neutrinos is a complex symmetric matrix of the  form

\begin{equation}
	M_{N} =
	\begin{pmatrix}
		k_1  Y_{1}^{(2)} + 2 Y_{1}^2 k_2 & Y_{3}^2(-k_2 ) & Y_{2}^2 (-k_2 ) \\
		Y_{3}^2 (-k_2 ) & 2 Y_{2}^2 k_2 & k_1  Y_{1}^{(2)} + Y_{1}^2 (-k_2 ) \\
		Y_{2}^2 (-k_2 ) & k_1  Y_{1}^{(2)} + Y_{1}^2 (-k_2 ) & 2 Y_{3}^2 k_2
	\end{pmatrix}.
	\label{eq:ME}
\end{equation}
This matrix structure exhibits special features for particular choices of the coupling constants, which will be discussed in Section \ref{sec3}. The matrix $M_{N}$ can be diagonalized as
\begin{equation}
U^{T}_R M_{N} U_R = \mathrm{diag}(m_{N_1}, m_{N_2},m_{N_3}),    
\end{equation}
where $U_R$  is the unitary diagonalizing matrix and $ m_{N_\alpha}$ ($ \alpha = 1,2,3$) are the masses of the three right-handed neutrinos.
The scalar potential of the model is
\begin{align}
V(H,\eta)
&= -m_H^{2}\,H^\dagger H +
m_\eta^{2}\,\eta^\dagger \eta + \lambda_{HH}\,(H^\dagger H )^2+\lambda_{\eta \eta}\,(\eta^\dagger \eta)^2
+\lambda_{H\eta}\,(H^\dagger H)\,(\eta^\dagger \eta)
\nonumber\\
&\quad 
+\lambda'_{H\eta}\,|H\eta|^{2}+\tilde{\lambda}\,Y^{(-2)}\,(H\eta)^2
+\text{h.c.}.
\end{align}
We define $\lambda \equiv \tilde{\lambda}\, Y^{(-2)}$, which is complex in general but can be considered real without loss of generality.

The inert scalar doublet $\eta$, consisting of a neutral real component $\eta_R$, 
a neutral imaginary component $\eta_I$ and a charged component $\eta^-$ with 
electric charge $-1$, can be written as
\begin{equation}
\eta =
\begin{pmatrix}
\dfrac{1}{\sqrt{2}}(\eta_R + i \eta_I) \\
\eta^-
\end{pmatrix}.
\end{equation}
After electroweak symmetry breaking, the Standard Model Higgs field acquires a vacuum expectation value $ \langle H \rangle = v/\sqrt{2} $ with $ v = 246\,\text{GeV} $, and the masses of the inert scalar bosons can then be written as 
\begin{align}
m^2_{\eta_{R}} &= m_\eta^2 + \frac{v^2}{2}
\left(\lambda_{H\eta} + \lambda'_{H\eta} + \, 2\lambda \right), \\
m^2_{\eta_{I}} &= m_\eta^2 + \frac{v^2}{2}
\left(\lambda_{H\eta} + \lambda'_{H\eta} - \, 2\lambda \right), \\
m^2_{\eta^-} &= m_\eta^2 + \frac{v^2}{2}\lambda_{H\eta}.
\end{align}
The mass-squared difference between the real and imaginary components of the inert scalar doublet, $\eta_R$ and $\eta_I$, respectively, is given by
\begin{equation}
\Delta m^2= m^2_{\eta_R} - m^2_{\eta_I} = 2\lambda v^2 .
\end{equation}
This mass splitting plays a crucial role in the model to satisfy neutrino phenomenology and achieve successful baryogenesis.

% Interestingly, in this scenario the splitting also depends on the value of $\tau$ and therefore cannot be chosen arbitrarily.
% For simplicity, we assume $m_{\eta_R} \simeq m_{\eta^-}$.

The charged lepton mass matrix can be written as 
\begin{equation}
	M_l =\frac{v}{\sqrt{2}}
\begin{pmatrix}
		\beta_e Y_{1}^0 & \beta_\mu Y_{3}^0 & \beta_\tau  Y_{2}^0 \\
		\beta_e Y_{3}^0 & \beta_\mu Y_{2}^0 & \beta_\tau Y_{1}^0 \\
		\beta_e Y_{2}^0 & \beta_\mu Y_{1}^0 & \beta_\tau Y_{3}^0
	\end{pmatrix}.
	\label{eq:Yc}
\end{equation}

The charged-lepton masses are obtained by diagonalizing the mass matrix
\( M_l \) with bi-unitary transformations \( V_L \) and \( V_R \), \textit{i.e.},
\begin{equation}
V_L^\dagger M_l V_R = \mathrm{diag}(m_e,\; m_\mu,\; m_\tau).
\end{equation}
Only the matrix \( V_L \) has physical effects. Numerically, \( V_L \) can be
determined by diagonalizing the Hermitian matrix $M_l^H = M_l M_l^\dagger$, which satisfies
\begin{equation}
V_L^\dagger M_l^H V_L = \mathrm{diag}(m_e^2,\; m_\mu^2,\; m_\tau^2).
\end{equation}
Since the charged-lepton masses are known with high precision, we can use
their experimental values to determine the three free parameters
$\beta_\alpha$ ($\alpha = e, \mu, \tau $) by means of these three relations
\begin{equation}\label{eq:cmass}
  \begin{aligned}
\begin{rcases}
 \mathrm{Tr}\!\left[M_l^H\right] &= |m_e|^2 + |m_\mu|^2 + |m_\tau|^2 , \\
\det\!\left[M_l^H\right] &= |m_e|^2 |m_\mu|^2 |m_\tau|^2 , \\
\left(\mathrm{Tr}\!\left[M_l^H\right]\right)^2
- \mathrm{Tr}\!\left[M_l^H\right]
&= 2\Big( |m_e|^2 |m_\mu|^2
+ |m_\mu|^2 |m_\tau|^2
+ |m_e|^2 |m_\tau|^2 \Big). 
\end{rcases}
\end{aligned}  
\end{equation}

\section{Nearly Degenerate Right-Handed Neutrino Mass Spectrum}\label{sec3}
The main subject of this paper which is tri-resonant leptogenesis needs nearly three degenerate right-handed neutrinos~\cite{Pilaftsis:2003gt,daSilva:2022mrx,Pilaftsis:2023snk}. This is a great challange from the model building prespective that we can have such a symmetry originated mass matrix where we can have such scenario. However we have identified that this is possible with existing A$_4$  non-holomorphic modular symmetry. As, we know that in RH neutrino mass matrix where  the RH neutrinos are A$_4$ triplets, we can have contraction under A$_4$ tensor product where contribution from the trivial singlet and symmetric part of the $\mathbf{3} \otimes \mathbf{3}$
 simultaneously arise in the mass matrix. The resultant matrix, as given in Eqn. (\ref{eq:ME}), can be written in the form 
\begin{equation}
\label{eq:M}
\tilde{M} =
\begin{pmatrix}
\tilde{M}_o+k d_1 & k d_4 & k d_5\\
k d_4 & k d_2 & \tilde{M}_o+k d_6\\
 k d_5 & \tilde{M}_o+k d_6 &  k d_3
\end{pmatrix}=\underbrace{ \begin{pmatrix}
\tilde{M}_o & 0 & 0\\
0 & 0 & \tilde{M}_o\\
0 & \tilde{M}_o & 0
\end{pmatrix}}_{\tilde{M}^{(0)}} +k \underbrace{ \begin{pmatrix}
d_1 & d_4 & d_5\\
d_4 & d_2 & d_6\\
d_5 & d_6 & d_3
\end{pmatrix}}_{\Delta},
\end{equation}
where $|k|\ll 1$. When $|k| = 0$, the matrix reduces to $\tilde{M}^{(0)}$.  
This matrix possesses an exact symmetry in the $(2,3)$ sector and can be diagonalized by the unitary matrix
\begin{equation}
\label{eq:U0}
U^{(0)} =
\begin{pmatrix}
0 & 1 & 0\\
i/\sqrt{2} & 0& 1/\sqrt{2}\\
-i/\sqrt{2} &0 & 1/\sqrt{2}
\end{pmatrix},
\end{equation}
such that $(U^{(0)})^T\tilde{M}^{(0)}U^{(0)}=\text{diag}(\tilde{M}_o,\tilde{M}_o,\tilde{M}_o)$, where $\tilde{M}_0$ denotes the eigenvalue of the matrix $\tilde{M}^{(0)}$. Therefore, the spectrum exhibits a threefold degeneracy.
 We can treat the  matrix elements of $\Delta$ in the Eqn. (\ref{eq:M}) as perturbation which can lift this degeneracy by softly breaking the  symmetry in (2,3) sector.
Therefore, the right-handed neutrino spectrum is naturally
nearly-degenerate, without the need for fine-tuning.
This feature is a direct consequence of  soft breaking of the symmetry in (2,3) sector in a way given in Eqns. (\ref{eq:ME}) or (\ref{eq:M}) and it provides a natural setting for tri-resonant or quasi-resonant leptogenesis.

\begin{table}[t]
\centering

\begin{tabular}{|l|c|c|c|}
\hline
\textbf{Parameter} & \textbf{Sampling Range}  &\textbf{NH Best Fit}  &\textbf{IH Best Fit}\\
\hline
$Re[\tau]$ & $\pm[0 \;-\; 0.5]$&$-8.872 \times 10^{-2}$&$-6.525 \times 10^{-3}$\\
$Im[\tau]$ & $[0.8 \;-\; 3]$&$1.231$&$1.336$\\
$k_{1}$ (GeV) & $[500 \;-\; 10^{4}]$&$3.557\times 10^{3}$&$3.467\times 10^{3}$\\
$k_{2}$ (GeV) &  $[10^{-9} \;-\; 10^{-2}]$&$6.191\times 10^{-7}$&$3.007\times 10^{-7}$\\
$\beta_{1}$& $\pm [10^{-6}- 10^{-3}]$&$-5.735\times 10^{-5}$&$-1.942\times 10^{-4}$\\
$\beta_{2}$& $\pm [10^{-6}- 10^{-3}]$&$-1.395\times 10^{-4}$&$2.037\times 10^{-4}$\\
$\beta_{3}$& $\pm [10^{-6}- 10^{-3}]$&$-1.681\times 10^{-4}$&$-2.557\times 10^{-5}$\\
 $m_{\eta}$ (GeV)& $[530-1500]$& $6.492\times 10^{2}$&$6.308\times 10^{2}$\\
 $\lambda$& $[10^{-5}-10^{-3}]$& $7.503 \times 10^{-4}$&$5.789\times 10^{-4}$\\
 % $\beta_{e}$& $-$& $-2.149\times 10^{-2}$&$-5.763\times 10^{-6}$\\
 % $\beta_{\mu}$& $-$& $-1.508\times 10^{-3}$&$1.807\times 10^{-2}$\\
 % $\beta_{\tau}$& $-$& $-7.849 \times 10^{-6}$&$1.150\times 10^{-3}$\\
 \hline
\end{tabular}
\caption{The input parameters and their numerical ranges used in the analysis (second column). 
The best-fit values of  free parameters for the normal hierarchy (NH) and inverted hierarchy (IH) are shown in the third and fourth columns, respectively. The minimum chi-square values are $\chi^2_{\min}(\mathrm{NH}) = 0.174$ and 
$\chi^2_{\min}(\mathrm{IH}) = 14.542$.}
\label{tab:params}
\end{table}

\section{Neutrino Phenomenology}\label{sec4}

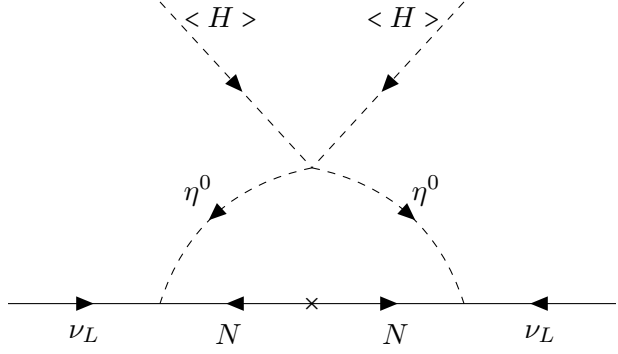
\begin{figure}[t]
    \centering
   \begin{tikzpicture}
\begin{feynman}
\vertex at (2,0) (i1);
\vertex at (-2,0) (i2);
\vertex at (0,0) (a);
\vertex at (0, 1.8) (d);
\vertex at (0,-2) (e);
\vertex at (4,0) (b);
\vertex at (-4,0) (c);
\vertex at (1.1,-0.4) () {\(N\)};
\vertex at (-1.1,-0.4) () {\(N\)};
\vertex at (2,4) (f);
\vertex at (-2,4) (g);
\vertex at (1.2,3.8) () {\(<H>\)};
\vertex at (-1.2,3.8) () {\(<H>\)};
\vertex at (1.5,1.5) () {\(\eta^{0}\)};
\vertex at (-1.5,1.5) () {\(\eta^{0}\)};
\vertex at (-3,-0.4) () {\(\nu_{L}\)};
\vertex at (3,-0.4) () {\(\nu_{L}\)};
\vertex at (0,0) () {\(\times\)};
\diagram*{
(a) -- [fermion] (i1), (a) -- [fermion] (i2),
(b) -- [fermion] (i1), (c) -- [fermion] (i2),
(d) -- [charged scalar, bend left] (i1), (d) -- [charged scalar,bend right]  (i2), (f) -- [charged scalar] (d),
 (g) -- [charged scalar] (d), 
};
\end{feynman}
\end{tikzpicture}
    \caption{One loop Feynman diagram for neutrino mass generation.}
    \label{fig:scofy}
\end{figure}

The neutrino mass arises from the one-loop diagram (Fig.~\ref{fig:scofy}) after spontaneous electroweak symmetry breaking. The resulting neutrino mass matrix is given by
% \begin{align}
% (M_\nu)_{ij} \simeq
% \sum_{\alpha=1}^{3}
% \frac{\tilde{Y}_{i\alpha}\, M{_\alpha}\, \tilde{Y}^{T}_{\alpha j}}{(4\pi)^2}
% \left[
% \frac{m_R^2}{m_R^2 - M_{\alpha}^2}
% \ln\!\left(\frac{m_R^2}{M_{\alpha}^2}\right)
% -
% \frac{m_I^2}{m_I^2 - M_{\alpha}^2}
% \ln\!\left(\frac{m_I^2}{M_{\alpha}^2}\right)
% \right].
% \end{align}
\begin{align}
M_\nu&=
\tilde{Y}\, m^D\, \tilde{Y}^{T}, 
\end{align}
where $m^D=\text{Diag}(m^D_1,m^D_2,m^D_3)$ with
\begin{align}
m^D_\alpha&=\frac{m_{N_\alpha}}{(4\pi)^2}\left[
\frac{m_{\eta_R}^2}{m_{\eta_R}^2 - m_{N_\alpha}^2}
\ln\!\left(\frac{m_{\eta_R}^2}{m_{N_\alpha}^2}\right)
-
\frac{m_{\eta_I}^2}{m_{\eta_I}^2 - m_{N_\alpha}^2}
\ln\!\left(\frac{m_{\eta_I}^2}{m_{N_\alpha}^2}\right)
\right],\;(\alpha=1,2,3),
\end{align}
and \( \tilde{Y} = Y_{\eta} U_R \) is the rotated Yukawa coupling in the basis where the right-handed neutrino mass matrix is diagonal. The neutrino mass matrix \( M_\nu \) can be diagonalized as
\begin{equation}
    U^T_\nu M_v U_\nu=\text{diag}(m_1,\,m_2,\,m_3)
\end{equation}
where \( m_i \) (\( i = 1,2,3 \)) are the neutrino masses. The Pontecorvo--Maki--Nakagawa--Sakata (PMNS) mixing matrix is given by $U = V_L^\dagger U_\nu$.
Before moving forward let us mention some relations used to calculate the mixing observables in the neutrino sector. Using mixing matrix $U$, the neutrino mixing angles can be obtained as
\begin{align}
\sin^2\theta_{13} &= |U_{13}|^2 ,  \\[4pt]
\sin^2\theta_{23} &= \frac{|U_{23}|^2}{1 - |U_{13}|^2} ,  \\[6pt]
\sin^2\theta_{12} &= \frac{|U_{12}|^2}{1 - |U_{13}|^2} . 
\end{align}
The Jarlskog CP invariant~\cite{Jarlskog:1985ht,RevModPhys.59.671,RevModPhys.60.575,RevModPhys.61.169,Krastev:1988yu}, which quantifies the CP violation arising from the Dirac phase $\delta_{CP}$, in the lepton sector is given by
\begin{align}
J_{CP}
&= \mathrm{Im}\!\left[
U_{11} U_{22} U_{12}^* U_{21}^*
\right]
= s_{23} c_{23} s_{12} c_{12} s_{13} c_{13}^2
\sin\delta_{CP},
\end{align}
where $s_{ij}\;(c_{ij})$ denotes $\sin\theta_{ij}\;(\cos \theta_{ij})$.
The CP invariants related to the Majorana phases $\alpha_{21}$ and $\alpha_{31}$ ~\cite{Nieves:1987pp,Aguilar-Saavedra:2000jom,Bilenky:2001rz,Nieves:2001fc} are
\begin{align}
I_1 &= \mathrm{Im}\!\left[ U_{11}^* U_{12} \right]
= c_{12} s_{12} c_{13}^2
\sin\!\left(\frac{\alpha_{21}}{2}\right), \\[6pt]
I_2 &= \mathrm{Im}\!\left[ U_{11}^* U_{13} \right]
= c_{12} s_{13} c_{13}
\sin\!\left(\frac{\alpha_{31}}{2} - \delta_{CP}\right).
\end{align}

The effective Majorana mass $m_{ee}$, which is currently being probed by neutrinoless double beta decay ($0\nu\beta\beta$) experiments and serves as a direct indicator of the Majorana nature of neutrinos, is given by
\begin{align}
m_{ee}
=
\left|
(U_{11})^2 m_1 + (U_{12})^2 m_2 + (U_{13})^2 m_3
\right|.
\tag{12}
\end{align}
An upper bound on the effective Majorana mass,
$m_{ee} < (28\text{--}122)\,\mathrm{meV}$ at $90\%$ confidence level,
has been reported by the KamLAND-Zen experiment~\cite{KamLAND-Zen:2024eml}, where the range reflects uncertainties in nuclear matrix element calculations.

\begin{table}[t]
\centering
\begin{tabular}{|c|c c|c c|}
\hline
\textbf{Parameter}
 & \multicolumn{2}{c|}{\textbf{NH (best fit)}} 
 & \multicolumn{2}{c|}{\textbf{IH ($\Delta\chi^2 = 5.9$)}} \\
 & bfp $\pm 1\sigma$ & $3\sigma$ range 
 & bfp $\pm 1\sigma$ & $3\sigma$ range \\
\hline
$\sin^2\theta_{12}$ 
& $0.3088^{+0.0067}_{-0.0066}$ 
& $0.2893 \rightarrow 0.3295$ 
& $0.3088^{+0.0067}_{-0.0066}$ 
& $0.2893 \rightarrow 0.3295$ \\

$\theta_{12}\,[^\circ]$ 
& $33.76^{+0.42}_{-0.41}$ 
& $32.54 \rightarrow 35.03$ 
& $33.76^{+0.42}_{-0.41}$ 
& $32.54 \rightarrow 35.03$ \\

$\sin^2\theta_{23}$ 
& $0.470^{+0.017}_{-0.014}$ 
& $0.435 \rightarrow 0.584$ 
& $0.550^{+0.013}_{-0.016}$ 
& $0.439 \rightarrow 0.584$ \\

$\theta_{23}\,[^\circ]$ 
& $43.29^{+0.96}_{-0.79}$ 
& $41.27 \rightarrow 49.86$ 
& $47.90^{+0.73}_{-0.92}$ 
& $41.51 \rightarrow 49.83$ \\

$\sin^2\theta_{13}$ 
& $0.02248^{+0.00055}_{-0.00059}$ 
& $0.02064 \rightarrow 0.02418$ 
& $0.02262^{+0.00057}_{-0.00056}$ 
& $0.02093 \rightarrow 0.02441$ \\

$\theta_{13}\,[^\circ]$ 
& $8.62^{+0.11}_{-0.11}$ 
& $8.26 \rightarrow 8.95$ 
& $8.65^{+0.11}_{-0.11}$ 
& $8.32 \rightarrow 8.99$ \\

$\delta_{\rm CP}\,[^\circ]$ 
& $212^{+26}_{-36}$ 
& $125 \rightarrow 365$ 
& $274^{+22}_{-25}$ 
& $203 \rightarrow 335$ \\

$\Delta m^2_{21}\,[10^{-5}\,\text{eV}^2]$ 
& $7.537^{+0.094}_{-0.100}$ 
& $7.236 \rightarrow 7.823$ 
& $7.537^{+0.094}_{-0.100}$ 
& $7.236 \rightarrow 7.822$ \\

$\Delta m^2_{3\ell}\,[10^{-3}\,\text{eV}^2]$ 
& $+2.511^{+0.021}_{-0.020}$ 
& $+2.450 \rightarrow +2.576$ 
& $-2.483^{+0.020}_{-0.020}$ 
& $-2.547 \rightarrow -2.421$ \\
\hline
\end{tabular}
\caption{Neutrino oscillation parameters from NuFIT 6.1 (2025)~\cite{Esteban:2024eli}, \url{http://www.nu-fit.org}. Best-fit values (bfp) with $1\sigma$ errors and $3\sigma$ ranges are shown. 
For normal hierarchy (NH) we take $\ell = 1$, while for inverted hierarchy (IH) we take $\ell = 2$.}
\label{tab:nufit61}
\end{table}

\begin{table}[t]
\centering
\begin{tabular}{|l|c|c|l|l|l|}
\hline
\textbf{Observable} & \textbf{NH Best Fit} & \textbf{IH Best Fit}  & \textbf{Observable} & \textbf{NH Best Fit} &\textbf{IH Best Fit}  \\
\hline
$\theta_{12}~({}^\circ)$ & $33.81$& $33.79$& $I_{1}$ & $4.358\times 10^{-1}$&$-4.034\times 10^{-4}$\\
$\theta_{13}~({}^\circ)$ & $8.63$& $8.68$  & $I_{2}$& $9.748\times 10^{-2}$&$1.240\times 10^{-1}$\\
$\theta_{23}~({}^\circ)$ & $43.16$& $44.76$  & $\sum_i m_i~(\mathrm{meV})$& $6.144\times 10^{1}$&$1.122\times 10^{2}$\\
$\Delta m_{21}^2~(10^{-5}\mathrm{eV}^2)$& $7.523$& $7.534$& $m_1~(\mathrm{meV})$& $2.250$&$5.027\times 10^{1}$\\
$\Delta m_{3l}^2~(10^{-3}\mathrm{eV}^2)$& $2.518$& $-2.484$& $m_2~(\mathrm{meV})$& $8.961$&$5.101\times 10^{1}$\\
$\delta_{\mathrm{CP}}~({}^\circ)$ & $319.127$& $0.216$& $m_3~(\mathrm{meV})$& $5.023\times 10^{1}$&$1.090\times 10^{1}$\\
$\alpha_{21}~({}^\circ)$ & $149.321$& $359.898$& $m_{N_1}~(10^{2} \mathrm{GeV})$& $7.673566660$&$9.702325117$\\
$\alpha_{31}~({}^\circ)$ & $22.800$& $179.284$& $m_{N_2}~(10^{2} \mathrm{GeV})$& $7.673566667$&$9.702325120$\\
$m_{ee}~(\mathrm{meV})$ & $1.135$& $4.910\times 10^{1}$& $m_{N_3}~(10^{2} \mathrm{GeV})$& $7.673566682$&$9.702325128$\\
$J_{\mathrm{CP}}$ & $-2.214\times 10^{-2}$& $1.288\times 10^{-4}$& & &\\
\hline
\end{tabular}
\caption{The neutrino observables evaluated at the best-fit points.
The minimum chi-square values are $\chi^2_{\min}(\mathrm{NH}) = 0.174$ and
$\chi^2_{\min}(\mathrm{IH}) = 14.542$. For normal hierarchy (NH) we take $\ell = 1$, while for inverted hierarchy (IH) we take $\ell = 2$.}
\label{tab:bf_observables}
\end{table}
 
\subsection{Numerical Analysis and Results}\label{ssec4.1}
To study the neutrino phenomenology we have numerically solved the Eqn. (\ref{eq:cmass}) and obtained the charged lepton diagonalizing matrix $V_L$. To obtain the neutrino masses and $U_\nu$ we have varied the free parameters in the ranges as shown in the Table \ref{tab:params}.
It should be noted that our goal is to identify a region of parameter space that can simultaneously satisfy neutrino oscillation data, account for dark matter, and generate the observed baryon asymmetry of the Universe. In this work, we consider the lightest dark sector particle, $\eta_I$, with mass $m_{\eta_I} $, as the dark matter (DM) candidate. To ensure consistency with the experimental bound on the DM relic density, we vary $m_{\eta_I}\simeq m_{\eta}$ in the range $[530-1500]~\mathrm{GeV}$. For appropriate choices of the scalar couplings, this mass range yields the correct relic abundance of dark matter \cite{Hambye:2009pw,Barbieri:2006dq,LopezHonorez:2006gr,LopezHonorez:2010eeh,Dolle:2009fn,LopezHonorez:2010tb,Goudelis:2013uca,Krawczyk:2013jta,Klasen:2013jpa,Arhrib:2013ela,Ilnicka:2015jba,Garcia-Cely:2015khw,Diaz:2015pyv,Belyaev:2016lok,Borah:2017dfn,Eiteneuer:2017hoh,Ahriche:2017iar,Avila:2021mwg}. Therefore, we do not discuss the dark matter phenomenology further in this analysis. Since the study of leptogenesis in our model requires small Yukawa couplings in order to suppress washout effects, we exploit the inverse relation between the mass splitting $\Delta m^2$, which is controlled by the coupling $\lambda$, and the strength of the Yukawa couplings. Accordingly, we vary $\lambda$ in the range $[10^{-5}\text{--}10^{-3}]$. This choice allows us to sufficiently suppress the Yukawa coupling strength while still satisfying the neutrino oscillation data.
In order to efficiently scan the parameter space we have used \texttt{emcee} package in python \cite{foreman2013emcee} and minimized the $\chi^2$ function  defined as
\begin{equation}
\chi^2 = \sum_{i = 1}^5 \left( \frac{O_i - O_i^{bf}}{\sigma_i} \right)^2,
     \label{}
\end{equation}
where $O_i$ represents the predicted value of the observable, $O_i^{bf}$ is the corresponding best-fit value and $\sigma_i$ is the 1$\sigma$ error  from the neutrino oscillation data  shown in Table \ref{tab:nufit61}. We have considered five  observables for our $\chi^2$ function which are the two neutrino mass-squared differences  and three neutrino mixing angles. 
Also, we restrict the parameter space by imposing $|\tau| > 1$.

\begin{figure}[t!]
    \centering
    \includegraphics[scale=0.45]{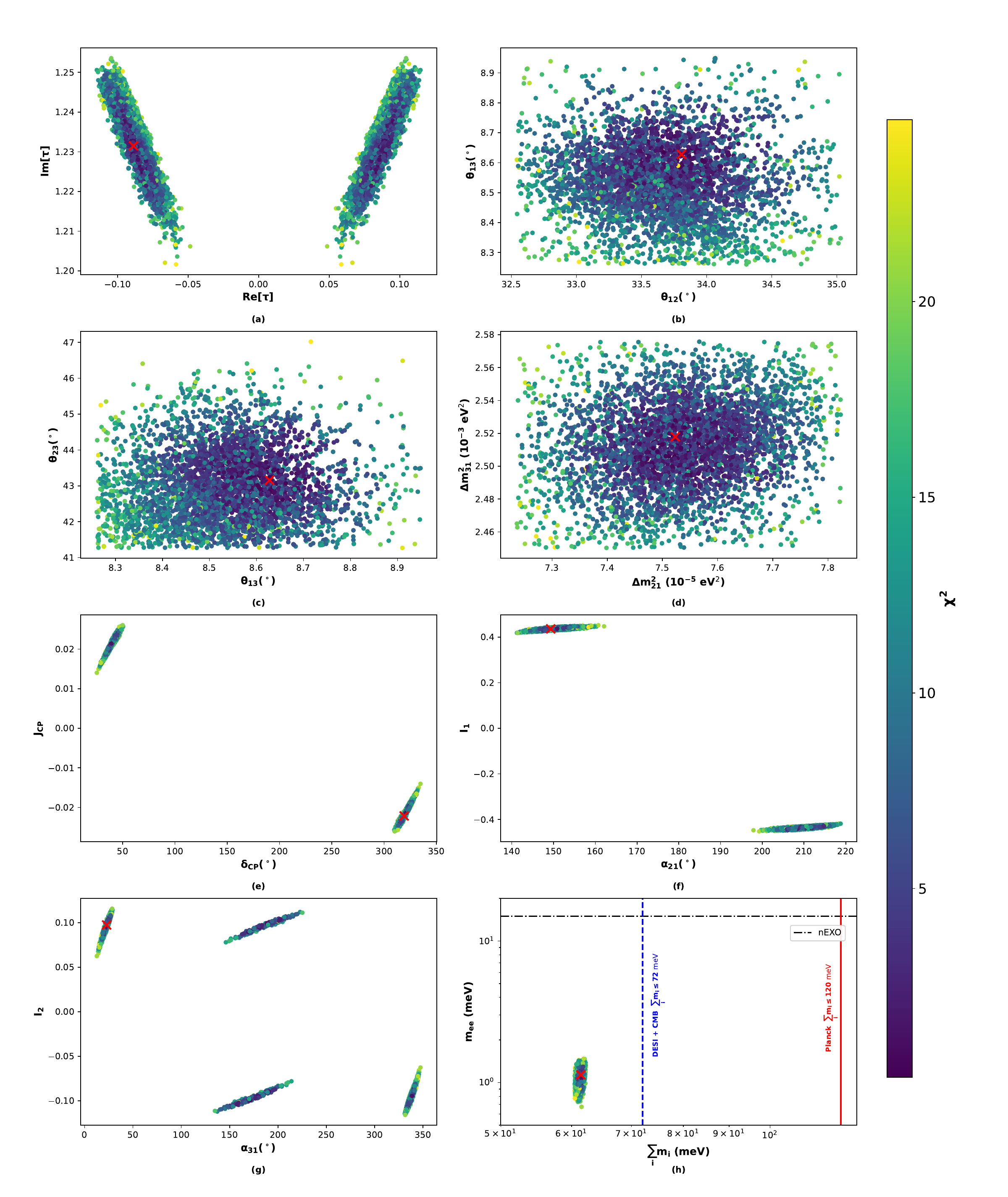}
    \caption{Model predictions for NH with neutrino oscillation parameters within the $3\sigma$ range of NuFIT~6.1 (Table \ref{tab:nufit61}). 
Point colors indicate the corresponding $\chi^2$ values, and the red cross (\textcolor{red}{\Large $\boldsymbol{\times}$}) symbol denotes the best-fit point with $\chi^2_{\min}=0.174$. The horizontal line in Fig.~\ref{fig:NH}(h) represents the $0\nu\beta\beta$ experimental sensitivity, and the vertical lines indicate cosmological upper bounds on $\sum_i m_i$.
}
    \label{fig:NH}
\end{figure}

\begin{figure}[t!]
    \centering
    \includegraphics[scale=0.45]{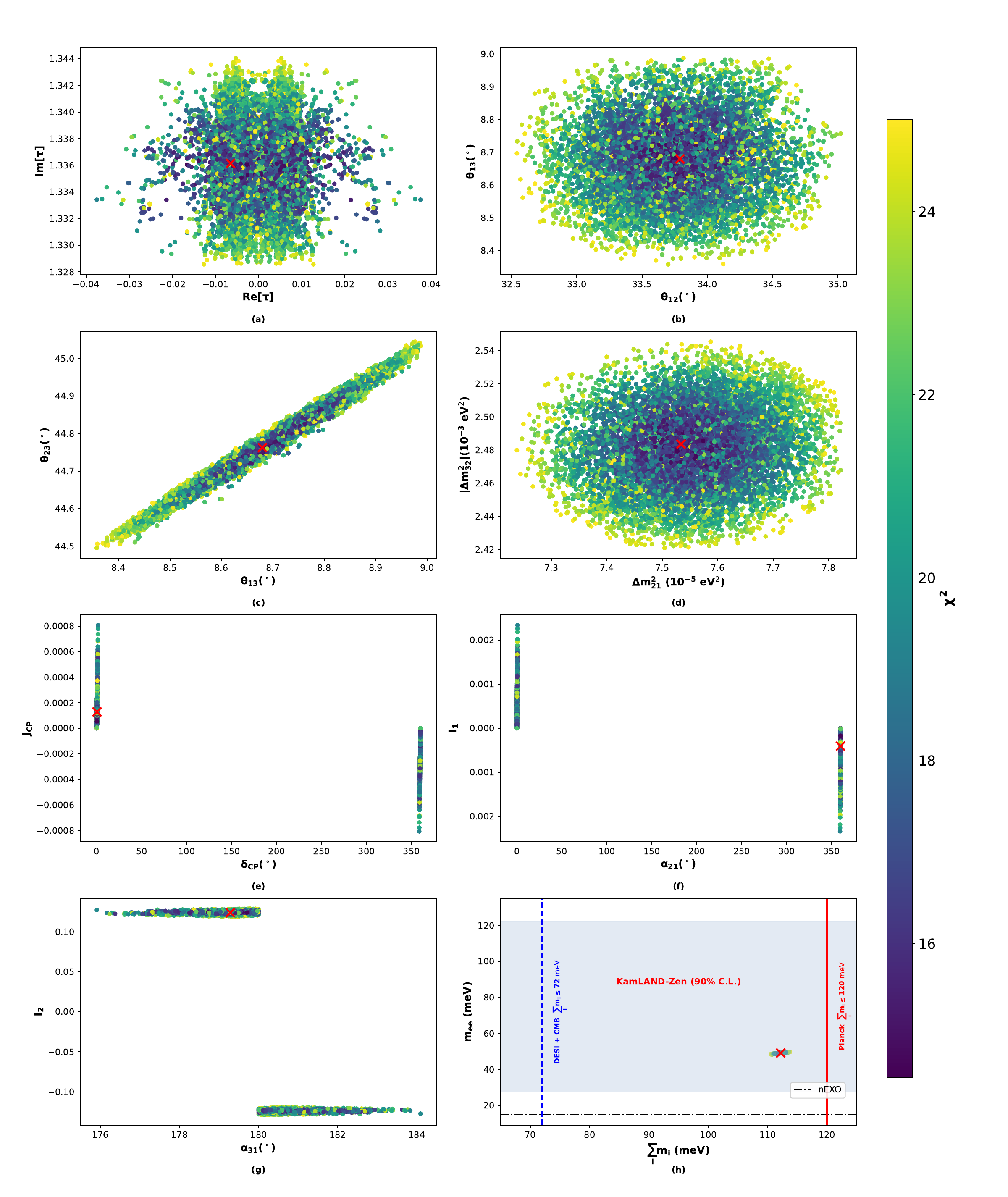}
    \caption{Model predictions for IH with neutrino oscillation parameters within the $3\sigma$ range of NuFIT~6.1 (Table \ref{tab:nufit61}). 
Point colors indicate the corresponding $\chi^2$ values, and the red cross (\textcolor{red}{\Large $\boldsymbol{\times}$}) symbol denotes the best-fit point with $\chi^2_{\min}=14.542$. The light blue shaded region in Fig.~\ref{fig:IH}(h) shows the KamLAND-Zen upper bound, the horizontal line represents the $0\nu\beta\beta$ experimental sensitivity, and the vertical lines indicate cosmological upper bounds on $\sum_i m_i$.}
    \label{fig:IH}
\end{figure}

The results for the normal hierarchy (NH) and inverted hierarchy (IH) of neutrino masses are shown in Fig.~\ref{fig:NH} and Fig.~\ref{fig:IH}, respectively, and lie within the $3\sigma$ range of the data presented in Table~\ref{tab:nufit61}.
The color coding denotes the corresponding $\chi^2$ values, while the red (\textcolor{red}{\Large $\boldsymbol{\times}$}) symbol indicates the global minimum, $\chi^2_{\min}=0.174$ for NH and $\chi^2_{\min}=14.542$ for IH. The corresponding best-fit parameter values are given in Table~\ref{tab:params}, and the observables are shown in Table~\ref{tab:bf_observables}.

The allowed regions for the modulus $\tau$ in the complex plane are displayed in Fig.~\ref{fig:NH}(a) (Fig.~\ref{fig:IH}(a)) for the NH (IH). For the NH, the imaginary part is constrained to a very narrow interval, $\mathrm{Im}[\tau] \in [1.20,\,1.25]$ while real part lies within $\mathrm{Re}[\tau] \in [\pm 0.05,\,\pm 0.11]$. In contrast, for the IH, one finds $\mathrm{Im}[\tau] \in [1.328,\,1.344]$ and $\mathrm{Re}[\tau] \in [0,\,\pm 0.04]$.
% Notably, most IH solutions cluster near the fixed point $i=\sqrt{-1}$. Since $\tau$ is the sole source of CP violation in the model, this behavior has direct implications for CP-violating observables.
As evident from Fig. \ref{fig:NH}(b) (\ref{fig:IH}(b)), there is no sharp correlation 
between $\theta_{12}-\theta_{13}$ in NH (IH). However, in contrast to 
Fig.~\ref{fig:NH}(c), the $\theta_{13}$-$\theta_{23}$ correlation is sharply 
constrained in the IH case (Fig. \ref{fig:IH}(c)). Moreover, the atmospheric 
mixing angle $\theta_{23}$ is predicted close to maximality $\theta_{23} \in [44.50^\circ,\,45.04^\circ]$, in the lower octant.
% The predicted neutrino mixing angles are shown in Figs.~\ref{fig:NH}(b)--\ref{fig:NH}(c) (Figs.~\ref{fig:IH}(b)--\ref{fig:IH}(c)) for the NH (IH). In the NH case, no sharp prediction emerges for the mixing angles. By contrast, in the IH scenario the atmospheric mixing angle is predicted to be close to maximality, $\theta_{23} \in [44.50^\circ,\,45.04^\circ]$, predominantly lying in the lower octant, $\theta_{23} < 45^\circ$ (Fig.~\ref{fig:IH}(c)).
The neutrino mass-squared differences are presented in Fig.~\ref{fig:NH}(d) (Fig.~\ref{fig:IH}(d)) for the NH (IH). The CP invariants and CP-violating phases are shown in Figs.~\ref{fig:NH}(e)--\ref{fig:NH}(g) (Figs.~\ref{fig:IH}(e)--\ref{fig:IH}(g)). In the NH case, we obtain sizable CP violation. In contrast, in the IH case the CP violation is relatively small, particularly from the Dirac phase $\delta_{CP}$ and the Majorana phase $\alpha_{21}$, which has important consequences for baryogenesis.
% In the NH case, where $\tau$ lies away from the fixed point, sizable CP violation is obtained. In contrast, the proximity of $\tau$ to $i$ in the IH case strongly suppresses CP violation, having important consequences for baryogenesis.
For the  NH, the Jarlskog invariant exhibits two localized regions:
$J_{\mathrm{CP}} \in [0.014,\,0.026]$ for $\delta_{\mathrm{CP}} \in [25^\circ,\,50^\circ]$, and
$J_{\mathrm{CP}} \in [-0.026,\,-0.014]$ for $\delta_{\mathrm{CP}} \in [310^\circ,\,335^\circ]$ (Fig.~\ref{fig:NH}(e)).
In contrast, for the IH, $J_{\mathrm{CP}}$ is highly suppressed, varying within
$J_{\mathrm{CP}} \in [0,\,0.0008]$ for $\delta_{\mathrm{CP}} \in [0^\circ,\,1.3^\circ]$, and
$J_{\mathrm{CP}} \in [-0.0008,\,0]$ for $\delta_{\mathrm{CP}} \in [358.7^\circ,\,360^\circ]$ (Fig.~\ref{fig:IH}(e)).
The Majorana CP invariant $I_1$ in the NH takes values
$I_1 \in [0.418,\,0.447]$ for $\alpha_{21} \in [141^\circ,\,162^\circ]$, and
$I_1 \in [-0.447,\,-0.418]$ for $\alpha_{21} \in [198^\circ,\,219^\circ]$ (Fig.~\ref{fig:NH}(f)).
For the IH, $I_1$ remains strongly suppressed, with
$I_1 \in [0,\,0.00234]$ for $\alpha_{21} \in [0^\circ,\,0.6^\circ]$ and
$I_1 \in [-0.00234,\,0]$ for $\alpha_{21} \in [359.4^\circ,\,360^\circ]$.
For the second Majorana CP invariant $I_2$ in the NH, four localized regions are obtained:
(i) $I_2 \in [0.078,\,0.112]$ for $\alpha_{31} \in [17^\circ,\,28^\circ]$;
(ii) $I_2 \in [0.063,\,0.114]$ for $\alpha_{31} \in [114^\circ,\,230^\circ]$;
(iii) $I_2 \in [-0.114,\,-0.063]$ for $\alpha_{31} \in [130^\circ,\,246^\circ]$;
(iv) $I_2 \in [-0.112,\,-0.078]$ for $\alpha_{31} \in [332^\circ,\,343^\circ]$ (Fig.~\ref{fig:NH}(g)).
In the IH case, $I_2$ is allowed only in a narrow region around
$I_2 \in [0.122,\,0.127]$ for $\alpha_{31} \in [176^\circ,\,180^\circ]$, and
$I_2 \in [-0.127,\,-0.122]$ for $\alpha_{31} \in [180^\circ,\,184^\circ]$ (Fig.~\ref{fig:IH}(g)).
The predicted ranges for the effective Majorana mass $m_{ee}$ and the sum of neutrino masses ($\Sigma_i m_i$) are shown in Fig.~\ref{fig:NH}(h) (Fig.~\ref{fig:IH}(h)) for the NH  (IH).
For the NH case, we have localized region around $m_{ee} \in [0.67,\,1.47]~\mathrm{meV}$, while the sum of neutrino masses varies within $\sum_i m_i \in [60.51,\,62.20]~\mathrm{meV}$. In this figure, the horizontal line represents the projected sensitivity of the nEXO~\cite{Licciardi:2017oqg} neutrinoless double beta ($0\nu\beta\beta$) decay experiment, while the vertical lines denote the cosmological upper bounds on the sum of neutrino masses from DESI+BAO ($\sum_i m_i \leq 72~\mathrm{meV}$)~\cite{DESI:2024mwx} and Planck ($\sum_i m_i \leq 120~\mathrm{meV}$)~\cite{Planck:2018vyg}. As can be seen, for the NH, $m_{ee}$ lies well below the nEXO sensitivity, whereas $\sum_i m_i \simeq 60.51$--$62.20~\mathrm{meV}$ is comfortably below the current cosmological upper limits (Fig.~\ref{fig:NH}(h)).
In contrast, for the IH, the effective Majorana mass lies in the range $m_{ee} \in [48.43,\,49.75]~\mathrm{meV}$, while the sum of neutrino masses spans $\sum_i m_i \in [110.52,\,113.72]~\mathrm{meV}$ (Fig.~\ref{fig:IH}(h)). In this figure, the light blue band indicates the upper bound on $m_{ee}$ obtained by KamLAND-Zen, namely $m_{ee} < (28$--$122)~\mathrm{meV}$ at $90\%$ C.L~\cite{KamLAND-Zen:2024eml}. In this case, $m_{ee}$ falls within the sensitivity reach of upcoming $0\nu\beta\beta$ experiments, while $\sum_i m_i$ is close to the Planck limit but significantly above the DESI+BAO upper bound. Given the KamLAND-Zen constraint, the inverted hierarchy scenario may be strongly constrained or potentially excluded by future experimental data. Similarly, forthcoming cosmological observations, particularly from Planck, are expected to further constrain $\sum_i m_i$ (Fig.~\ref{fig:IH}(h)). Finally, we briefly address lepton flavor violation (LFV). The Yukawa couplings in our framework are naturally small, leading to LFV rates well below existing experimental limits~\cite{MEG:2016leq}. Consequently, LFV constraints do not impose additional restrictions on the parameter space considered here, and we therefore omit a detailed discussion.

\section{Tri-Resonant Leptogenesis}\label{sec5}

In this section, we discuss \textit{tri-resonant leptogenesis}. As outlined in the introduction, resonant leptogenesis is one of the viable mechanisms for generating the observed baryon asymmetry of the Universe at energy scales below $1~\mathrm{TeV}$ within the scotogenic model. In this framework, accommodating neutrino oscillation data typically requires relatively large Yukawa couplings. However, successful baryogenesis demands smaller Yukawa couplings in order to suppress washout processes. Reducing the Yukawa couplings generally suppresses the CP asymmetry, thereby making baryogenesis difficult to achieve.
This situation changes significantly in the presence of  nearly degenerate right-handed (RH) neutrinos. When the mass splitting between two RH neutrinos, defined as $\delta_{ij} = |m_{N_i} - m_{N_j}|$, becomes comparable to half of the decay width of the decaying particle, $\Gamma_{i,j}$, \textit{i.e.},
\begin{equation}
\delta_{ij} \simeq \frac{\Gamma_{i,j}}{2},
\end{equation}
the CP asymmetry undergoes resonant enhancement~\cite{Pilaftsis:1997jf,Pilaftsis:2003gt,Pilaftsis:2005rv}. As a result, large CP asymmetries can be generated even for relatively small Yukawa couplings, allowing for successful baryogenesis.

In our framework, we have three nearly degenerate RH neutrinos with a mass ordering $m_{N_1} \lesssim m_{N_2} \lesssim m_{N_3}$. In this case, the enhancement is further amplified since all three RH neutrinos can contribute to the generation of the lepton asymmetry through their decays. Moreover, as we study leptogenesis at low energy scales (below $1~\mathrm{TeV}$), flavor effects become relevant and can further enhance the CP asymmetry. Consequently, we anticipate that successful baryogenesis can be achieved in our model even in the deep strong-washout regime, as will be demonstrated later in this section.

In the tri-resonant leptogenesis scenario, a maximally enhanced CP asymmetry is generated due to the mixing among all three right-handed singlet neutrinos~\cite{Pilaftsis:2003gt,daSilva:2022mrx,Pilaftsis:2023snk}. The full set of resonant effects contributing to the CP asymmetry arising from this mixing can be systematically captured through the resummed effective Yukawa couplings $(Y_+)_{l\alpha}$ and $(Y_-)_{l\alpha}$. The effective Yukawa coupling $(Y_+)_{l\alpha}$ is associated to the $L \eta N$ interaction, which can be written as

\begin{equation}
\begin{aligned}
\left(Y_+\right)_{l\alpha}
&= Y_{l\alpha}+iV_{l\alpha}
- i \sum_{\beta,\gamma=1}^{3}
\left| \epsilon_{\alpha\beta\gamma} \right|
\, Y_{l\beta}
\\[2mm]
&\quad \times
\frac{
m_{N_\alpha}\left(M_{\alpha\alpha\beta}+M_{\beta\beta\alpha}\right)
- i R_{\alpha\gamma}
\left[
M_{\alpha\gamma\beta}\left(M_{\alpha\alpha\gamma}+M_{\gamma\gamma\alpha}\right)
+ M_{\beta\beta\gamma}\left(M_{\alpha\gamma\alpha}+M_{\gamma\alpha\gamma}\right)
\right]
}{
m_{N_\alpha}^2 - m_{N_\beta}^2
+ 2 i m_{N_\alpha}^2 A_{\beta\beta}
+ 2 i\, \mathrm{Im}\, R_{\alpha\gamma}
\left(
m_{N_\alpha}^2 \left| A_{\beta\gamma} \right|^2
+ m_{N_\beta} m_{N_\gamma} \mathrm{Re}\, A_{\beta\gamma}^2
\right)
},
\end{aligned}
\label{eq:effY}
\end{equation}

where $Y=V_L^\dagger Y_\eta U_R$, $\epsilon_{\alpha\beta\gamma}$ is the anti-symmetric Levi--Civita symbol, $M_{\alpha\beta\gamma} \equiv m_{N_\alpha} A_{\beta\gamma}$ and
\begin{equation}
A_{\alpha\beta}
=
\sum_{l=1}^{3}
\frac{Y_{l\alpha}\, Y_{l\beta}^{ *}}{16\pi}
=
\frac{\left(Y^{\dagger} Y \right)^{*}_{\alpha\beta}}{16\pi}
\, ,
\end{equation}
\begin{equation}
V_{l\alpha}
=
\sum_{k=1}^{3}\;
\sum_{\gamma \neq \alpha}
\frac{Y^{ *}_{k\alpha}\, Y_{k\gamma}\, Y_{l\gamma}}
{16\pi}
\, f\!\left(\frac{m_{N_\gamma}^2}{m_{N_\alpha}^2}\right),
\end{equation}
\begin{equation}
R_{\alpha\beta}
\equiv
\frac{m_{N_\alpha}^2}
{m_{N_\alpha}^2 - m_{N_\beta}^2 + 2 i m_{N_\alpha}^2 A_{\beta\beta}} \, .
\label{eq:4.4}
\end{equation}
Here, $A_{\alpha \beta}$ and $V_{l\alpha}$  represent the absorptive transition amplitudes arising from the propagator and vertex contributions, respectively. The quantity
$f(x)=\sqrt{x}\left[1-(1+x)\ln\!\left(\frac{1+x}{x}\right)\right]$
denotes the Fukugita–Yanagida one-loop function~\cite{Fukugita:1986hr}.

The  effective Yukawa coupling $(Y_-)_{l\alpha}$ associated to $\overline{L} \overline{\eta} N$ interaction can be obtained from Eqn. (\ref{eq:effY}) by replacing $Y_{l\alpha}$ with $Y_{l\alpha}^*$. The radiatively corrected  partial decay  widths in terms of these effective Yukawa couplings can be written as
 %  Decay Width
\begin{align}
\Gamma_\alpha
&=
\sum_{l=e,\mu,\tau} \Gamma(N_\alpha \to L_l + \eta)
=
\frac{m_{N_\alpha}}{16\pi}
(Y_{+}^\dagger Y_{+})_{{\alpha \alpha}}\Big(1-\frac{m_\eta^2}{m_{N_\alpha}^2} \Big)^2,
\\[1ex]
\overline{\Gamma}_\alpha
&=
\sum_{l=e,\mu,\tau} \Gamma(N_\alpha \to \overline{L}_l +\overline{\eta})
=
\frac{m_{N_\alpha}}{16\pi}
(Y_{-}^\dagger Y_{-})_{\alpha \alpha} \Big(1-\frac{m_\eta^2}{m_{N_\alpha}^2} \Big)^2,
\end{align}
and we can write total decay width as
\begin{equation}
    \Gamma^T_\alpha=\Gamma_\alpha + \overline{\Gamma}_\alpha=\frac{m_{N_\alpha}}{16\pi}\Big[(Y_{+}^\dagger Y_{+})_{\alpha \alpha} +
(Y_{-}^\dagger Y_{-})_{\alpha \alpha}\Big]\Big(1-\frac{m_\eta^2}{m_{N_\alpha}^2} \Big)^2.
\end{equation}
The CP-asymmetry matrix element $\varepsilon^{\alpha}_{ll'}$ can be written as the ratio of the difference of these partial decay widths to the total deacy width, and it determines the asymmetry generated in each $N_\alpha$ decay as 
%CP- asymmetry matrix
\begin{align}
\varepsilon^{\alpha}_{ll^\prime}
=
\frac{
\overline{P}^{\alpha}_{ll^\prime}\,\overline{\Gamma}_\alpha
-
P^{\alpha}_{ll^\prime}\,\Gamma_\alpha
}{
\Gamma_\alpha^T
},
\end{align} 
The two projectors $P^{\alpha}_{ll^\prime}$ and $\overline{P}^{\alpha}_{ll^\prime}$ in the above expression are given  by
\begin{align}
P^{\alpha}_{ll^\prime}
= C_{\alpha l} C^*_{\alpha l^\prime },
\qquad
\overline{P}^{\alpha}_{ll^\prime}
= \overline{C}_{\alpha l} \overline{C}^*_{\alpha l^\prime },
\end{align}
with amplitudes 
\begin{align}
C_{\alpha l}
=
\frac{(Y_{+})_{l \alpha}}
{\sqrt{(Y_{+}^\dagger Y_{+})_{\alpha \alpha}}},
\qquad
\overline{C}_{\alpha l}
=
\frac{(Y_{-})_{l \alpha}}
{\sqrt{(Y_{-}^\dagger Y_{-})_{\alpha \alpha}}}.
\end{align}
 At tree level, these amplitudes becomes
\begin{equation}
    C^0_{\alpha l}
=
\frac{Y_{l \alpha}}
{\sqrt{(Y^\dagger Y)_{\alpha \alpha}}}.
\end{equation}

In the CP asymmetry matrix, each diagonal entry corresponds to the flavored CP asymmetry $\varepsilon^\alpha_{ll} = \varepsilon_{\alpha l}$, which is real. In contrast, the off-diagonal entries satisfy
$\varepsilon^\alpha_{l l'} = (\varepsilon^\alpha_{l' l})^*$ and can, in general, be complex. The total CP asymmetry $\varepsilon_\alpha$ in the decay of $N_\alpha$ is obtained by taking the trace of the CP asymmetry matrix, i.e., $\varepsilon_\alpha = \sum\limits_{l} \varepsilon^\alpha_{ll}.$ The total CP asymmetry then can be written as $\varepsilon_T=\sum \limits_{\alpha} \varepsilon_\alpha$.\\
To determine the evolution of the right-handed (RH) neutrino number density 
$n_{N_\alpha}$ and the lepton asymmetry number density $n_{\ell\ell'}$, 
generated in each flavor through the decays of $N_\alpha$, we must solve 
evolution equations that explicitly account for flavor effects. The evolution 
is studied as a function of the dimensionless parameter
$z \equiv m_{N_1}/T$, where $m_{N_1}$ is the mass of the lightest RH neutrino and $T$ is the temperature of the Universe.  For this purpose, we employ density matrix equations, which are particularly well suited for tracking flavor dynamics and can be written as~\cite{Abada:2006fw,Barbieri:1999ma,DeSimone:2006nrs,Blanchet:2006ch,Blanchet:2011xq,Moffat:2018wke,Granelli:2021fyc,Liu:2024eki}
\begin{align}\label{eq:denmax}
  \frac{d n_{N_\alpha}}{dz}
&=
- D_\alpha
\left(
n_{N_\alpha} - n_{N_\alpha}^{\rm eq}
\right),  \notag \\ 
%%%%%%%%%%% ee
\frac{d n_{ee}}{dz}
&= \sum_{\alpha=1}^{3}
\Bigg\{
\varepsilon^{\alpha}_{ee}\,
D_\alpha
\left(
n_{N_\alpha} - n_{N_\alpha}^{\rm eq}
\right)
-
W_\alpha
\Big[
|C^0_{\alpha e}|^2\, n_{ee}
+ \mathrm{Re}
\left[
C^{0*}_{\alpha e} \Big(C^0_{\alpha\mu}\, n^*_{\mu e}
+  C^0_{\alpha\tau}\, n^*_{\tau e}\Big)
\right]
\Big]
\Bigg\},
\notag \\
%%%%%%%%%%%%%%%%%%% uu
\frac{d n_{\mu\mu}}{dz}
&=
\sum_{\alpha=1}^{3}
\Bigg\{
\varepsilon^{\alpha}_{\mu\mu}\,
D_\alpha
\left(
n_{N_\alpha} - n_{N_\alpha}^{\rm eq}
\right)
-
W_\alpha
\Big[
|C^0_{\alpha\mu}|^2\, n_{\mu\mu}
+ \mathrm{Re}
\left[
C^{0*}_{\alpha\mu} \Big(C^0_{\alpha e}\, n_{\mu e}
+ C^0_{\alpha\tau}\, n^*_{\tau\mu}\Big)
\right]
\Big]
\Bigg\},
\notag \\
%%%%%%%%%%%%%%%%%%%%%%%%%%%%%%%%%% tt
\frac{d n_{\tau\tau}}{dz}
&=
\sum_{\alpha=1}^{3}
\Bigg\{
\varepsilon^{\alpha}_{\tau\tau}\,
D_\alpha
\left(
n_{N_\alpha} - n_{N_\alpha}^{\rm eq}
\right)
-
W_\alpha
\Big[
|C^0_{\alpha\tau}|^2\, n_{\tau\tau}
+ \mathrm{Re}
\left[
C^{0*}_{\alpha\tau} \Big( C^0_{\alpha e}\, n_{\tau e}
+  C^0_{\alpha\mu}\, n_{\tau\mu} \Big)
\right]
\Big]
\Bigg\},
\notag \\
%%%%%%%%%%%%%%%%%%%%%%%%%%%%%%
%%%%%%%%%%%%%%%%%%%%%%%%%%%%%%%%%%%%% tu
\frac{d n_{\tau\mu}}{dz}
&=
\sum_{\alpha=1}^{3}
\Bigg\{
\varepsilon^{\alpha}_{\tau\mu}\,
D_\alpha
\left(
n_{N_\alpha} - n_{N_\alpha}^{\rm eq}
\right)
-
\frac{1}{2} W_\alpha
\Big[
n_{\tau\mu}
\big(
|C^0_{\alpha\tau}|^2 + |C^0_{\alpha\mu}|^2
\big)+ C^{0*}_{\alpha\mu} C^0_{\alpha\tau}
\Big(
n_{\tau\tau} + n_{\mu\mu}
\Big)
\notag \\
&\qquad\qquad
+ C^{0*}_{\alpha e} C^0_{\alpha\tau}\, n^*_{\mu e}
+ C^{0*}_{\alpha\mu} C^0_{\alpha e}\, n_{\tau e}
\Big]
\Bigg\}
-
\Big[
\mathrm{Im}(\Lambda_\tau)
+ \mathrm{Im}(\Lambda_\mu)
\Big]
\frac{n_{\tau\mu}}{Hz}\,
,
\notag \\
%%%%%%%%%%%%%%%%%%%%%%%%%%%%%%%%%%%%%%%%%%
%%%%%%%%%%%%%%%% te
\frac{d n_{\tau e}}{dz}
&=
\sum_{\alpha=1}^{3}
\Bigg\{
\varepsilon^{\alpha}_{\tau e}\,
D_\alpha
\left(
n_{N_\alpha} - n_{N_\alpha}^{\rm eq}
\right)
-
\frac{1}{2} W_\alpha
\Big[
n_{\tau e}
\big(
|C^0_{\alpha e}|^2 + |C^0_{\alpha\tau}|^2
\big)+ C^{0*}_{\alpha e} C^0_{\alpha\tau}
\Big(
n_{ee} + n_{\tau\tau}
\Big) \notag
\\
&\qquad\qquad
+ C^{0*}_{\alpha\mu} C^0_{\alpha\tau}\, n_{\mu e}
+ C^{0*}_{\alpha e} C^0_{\alpha\mu}\, n_{\tau\mu}
\Big]
\Bigg\}
-
\mathrm{Im}(\Lambda_\tau)
\frac{n_{\tau e}}{Hz}\, ,
\notag \\
%%%%%%%%%%%%%%%%%%%%%%%%%%%%%
%%%%%%%%%%%%%%%%%%%%%%%%%%%  ue
\frac{d n_{\mu e}}{dz}
&=
\sum_{\alpha=1}^{3}
\Bigg\{
\varepsilon^{\alpha}_{\mu e}\,
D_\alpha
\left(
n_{N_\alpha} - n_{N_\alpha}^{\rm eq}
\right)
-
\frac{1}{2} W_\alpha
\Big(
n_{\mu e}
\big(
|C^0_{\alpha e}|^2 + |C^0_{\alpha\mu}|^2
\big)
+ C^{0*}_{\alpha e} C^0_{\alpha\mu}
\Big(
n_{ee} + n_{\mu\mu}
\Big) \notag
\\ & \qquad \qquad
+ C^{0*}_{\alpha e} C^0_{\alpha\tau}\, n^*_{\tau\mu}
+ C^{0*}_{\alpha\tau} C^0_{\alpha\mu}\, n_{\tau e}
\Big)
\Bigg\}-
\mathrm{Im}(\Lambda_\mu)
\frac{n_{\mu e}}{Hz}  \, .
\end{align}
In these equations $n_{N_\alpha}^{eq}$ is the equilibrium number density of the RH neutrino $N_\alpha\;(\alpha=1,2,3)$ and symbol $D_\alpha$ and $W_\alpha$ denote the decays and washouts\footnote{As, we are in strong-washout region so we can safely neglect the $\Delta L=1$ and $\Delta L=2$ scattering processes~\cite{Davidson:2008bu,Hugle:2018qbw}.} due to inverse decay, respectively, in thermal plasma and can be written as
%  Washouts
\begin{align}
D_\alpha(z)
&=
K_\alpha\, x_\alpha\, z\,
\frac{\mathcal{K}_1(z_\alpha)}{\mathcal{K}_2(z_\alpha)},
\\[1ex]
W_\alpha(z)
&=
\frac{1}{4}K_\alpha \sqrt{x_\alpha}\, z_\alpha^3\, \mathcal{K}_1(z_\alpha).
\end{align}
In the above expressions, $K_\alpha = \Gamma^T_\alpha / H(z=1)$ is the decay 
parameter, where the Hubble expansion rate is given by
\begin{equation}
H(z) = \sqrt{\frac{4\pi^3 g_\star}{45}}\,\frac{m_{N_1}^2}{z^2 M_{\mathrm{Pl}}}.
\end{equation}
Here $g_\star = 116$ denotes the effective number of relativistic degrees of 
freedom in the early Universe, and 
$M_{\mathrm{Pl}} = 1.22 \times 10^{19}\,\mathrm{GeV}$ is the Planck mass. 
The symbols $\mathcal{K}_1$ and $\mathcal{K}_2$ denote the modified Bessel 
functions of the second kind and we  define $x_\alpha \equiv m_{N_\alpha}^2/m_{N_1}^2, \; z_\alpha \equiv z \sqrt{x_\alpha}$. The symbol $\Lambda_l$ in Eq.~\eqref{eq:denmax} denotes the self-energy associated with the charged lepton of flavor $l$. Its imaginary part is related to the Yukawa coupling $y_l$ of the corresponding charged lepton and temperature $T$, 
and is approximately given by~\cite{Blanchet:2011xq}
\begin{equation}
\operatorname{Im}(\Lambda_l) \simeq 8 \times 10^{-3}\, y_l^2\, T.
\end{equation}
The lepton density matrix $n_{ll'}$ is a $3\times 3$ matrix whose diagonal 
elements $n_{ll}$ are real, while the off-diagonal elements are, in general, 
complex. The total lepton asymmetry can be obtained from this matrix by taking 
its trace,
\begin{equation}
n_{B-L}(z) = \sum_{l = e,\mu,\tau} n_{ll}(z).
\end{equation}
This asymmetry is subsequently partially converted into a baryon asymmetry 
through $B-L$ conserving sphaleron transitions. The final baryon asymmetry can 
be written as
\begin{equation}
\eta_B =0.012 \; n_{B-L}(z_{\rm sph}),
\end{equation}
where $z_{\rm sph} = m_{N_1} / T_{\rm sph}$, and 
$T_{\rm sph} = 131.7\,\mathrm{GeV}$ is the sphaleron freeze-out temperature~\cite{DOnofrio:2014rug}. This result is then compared with the experimentally observed value of the 
baryon asymmetry, $\eta_B = (6.12 \pm 0.04) ~\times~ 10^{-10}$, as inferred from measurements of the cosmic microwave background (CMB)~\cite{Planck:2018vyg}.

\begin{table}[t]
\centering
\begin{tabular}{|c|c|c|c|c|}
\hline
\textbf{Quantity} & \textbf{BP1 (NH)} & \textbf{BP2 (NH)} & \textbf{BP1 (IH)} & \textbf{BP2 (IH)} \\
\hline
$Re[\tau]$     &  $1.088\times 10^{-1}$&  $-7.598\times 10^{-2}$&  $-6.586\times 10^{-3}$&  $-9.428\times 10^{-3}$\\
$Im[\tau]$  &  $1.247$&  $1.214$&  $1.331$&  $1.333$\\
$k_1$ ($10^{3}$ GeV)&  $2.366$&  $2.922$&  $2.064$&  $2.005$\\
$k_2$ (GeV)&  $2.987\times 10^{-4}$&  $7.015 \times 10^{-7}$&  $6.626 \times 10^{-6}$&  $1.471 \times 10^{-6}$\\
$\beta_1$  &  $-3.868 \times 10^{-5}$&  $-7.487 \times 10^{-5}$&  $-1.985\times 10^{-4}$&  $-2.069\times 10^{-4}$\\
$\beta_2$  &  $-1.282\times 10^{-4}$&  $-1.437\times 10^{-4}$&  $2.082 \times 10^{-4}$&  $2.170 \times 10^{-4}$\\
$\beta_3$  &  $-1.566\times 10^{-4}$&  $-1.723 \times 10^{-4}$&  $-2.665 \times 10^{-5}$&  $-2.752 \times 10^{-5}$\\
$m_\eta$ ($10^{2}$ GeV) &  $5.313$&  $5.314$&  $5.665$&  $5.441$\\
$\lambda$  &  $7.626\times 10^{-4}$&  $5.762\times 10^{-4}$&  $5.495 \times 10^{-4}$&  $4.806 \times 10^{-4}$\\
$m_{N_1}$ ($10^{2}$ GeV) &  $5.372075$&  $5.92791349$&  $5.7227086$&  $5.57781227$\\
$m_{N_2}$ ($10^{2}$ GeV) &  $5.372078$&  $5.92791350$&  $5.7227087$&  $5.57781228$\\
$m_{N_3}$ ($10^{2}$ GeV) &  $5.372085$&  $5.92791352$&  $5.7227089$&  $5.57781232$\\
$|\varepsilon_T|$  &  $3.438 \times 10^{-4}$&  $6.723 \times 10^{-2}$&  $2.412\times 10^{-4}$&  $1.471\times 10^{-3}$\\
$K_1$  &  $1.211 \times 10^{3}$&  $1.098\times 10^{5}$&  $2.613\times 10^{3}$&  $1.696\times 10^{4}$\\
$K_2$  &  $1.249\times 10^{3}$&  $1.146\times 10^{5}$&  $6.172\times 10^{2}$&  $4.038\times 10^{3}$\\
$K_3$  &  $1.838\times 10^{3}$&  $1.765\times 10^{5}$&  $2.653\times 10^{3}$&  $1.722\times 10^{4}$\\
$\delta _{12}/m_{N_1}$  &  $5.940 \times 10^{-7}$&  $1.407\times 10^{-9}$&  $1.046\times 10^{-8}$&  $2.371 \times 10^{-9}$\\
$\delta _{23}/m_{N_2}$&  $1.408 \times 10^{-6}$&  $2.945 \times 10^{-9}$&  $3.008\times 10^{-8}$&  $6.855 \times 10^{-9}$\\
$\delta_{13}/m_{N_1}$&  $2.002\times 10^{-6}$&  $4.351 \times 10^{-9}$&  $4.054 \times 10^{-8}$&  $9.226 \times 10^{-9}$\\
$\eta_B$  &  $6.12\times 10^{-10}$&  $6.12 \times 10^{-10}$&  $6.12 \times 10^{-10}$&  $6.12 \times 10^{-10}$\\
$\chi^2$  &  $15.593$&  $9.792$&  $20.203$&  $17.665$\\
\hline
\end{tabular}
\caption{The benchmark points (BP1 and BP2) for both NH and IH considered in the numerical analysis of the density matrix equations.}
\label{tab:BP}
\end{table}

\begin{figure}[t]
    \centering
    \includegraphics[scale=0.5]{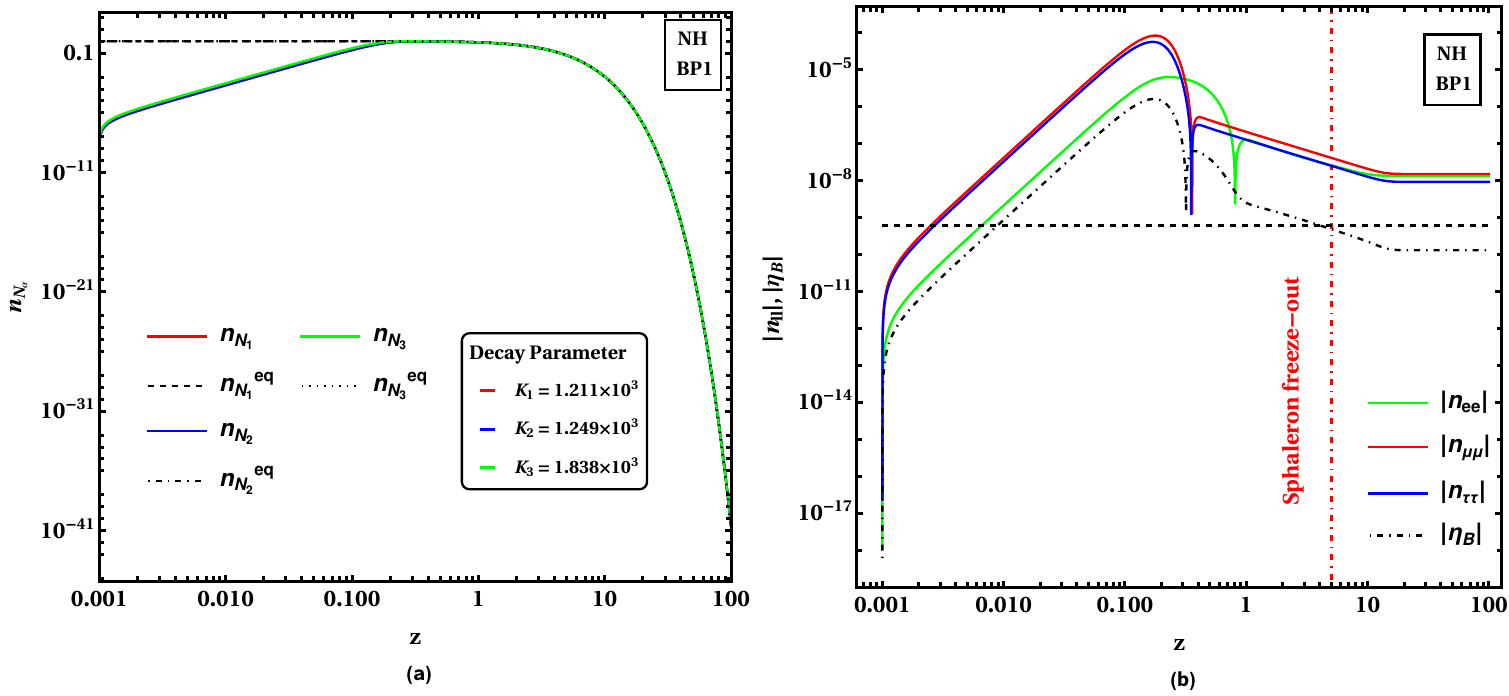}\\
    \includegraphics[scale=0.5]{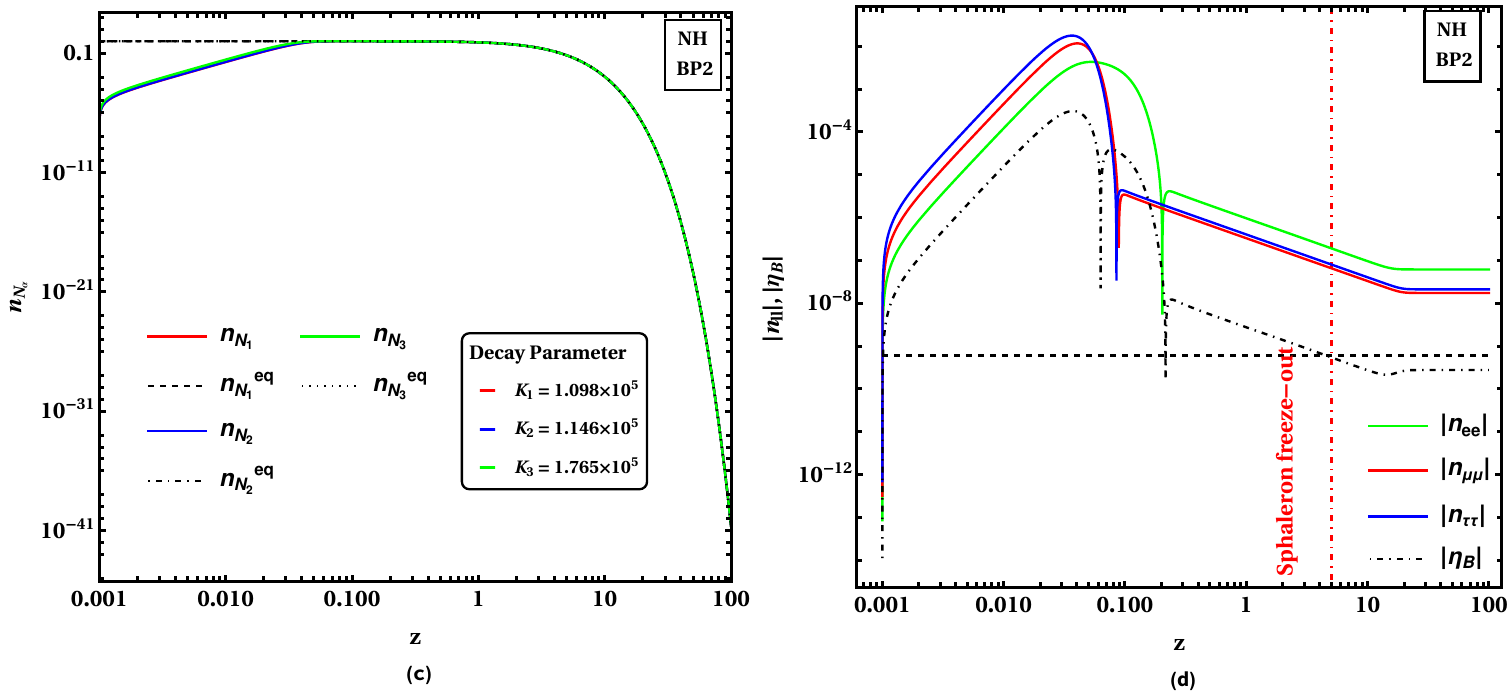}
    \caption{Evolution of the number density of each right-handed neutrino in 
Fig.~\ref{fig:NHB}(a) (\ref{fig:NHB}(c)) and evolution of the number 
density of each lepton flavor together with the baryon asymmetry in 
Fig.~\ref{fig:NHB}(b) (\ref{fig:NHB}(d)) as a function of $z$ for BP1 (BP2) 
in the normal hierarchy (NH) case. The horizontal dashed black line and 
the vertical dot-dashed red line in Fig.~\ref{fig:NHB}(b) and 
Fig.~\ref{fig:NHB}(d) indicate the experimentally observed value of the 
baryon asymmetry and the sphaleron freeze-out, respectively. Zero initial 
abundance is assumed for all particle species.
}
    \label{fig:NHB}
\end{figure}

\begin{figure}[t]
    \centering
    \includegraphics[scale=0.5]{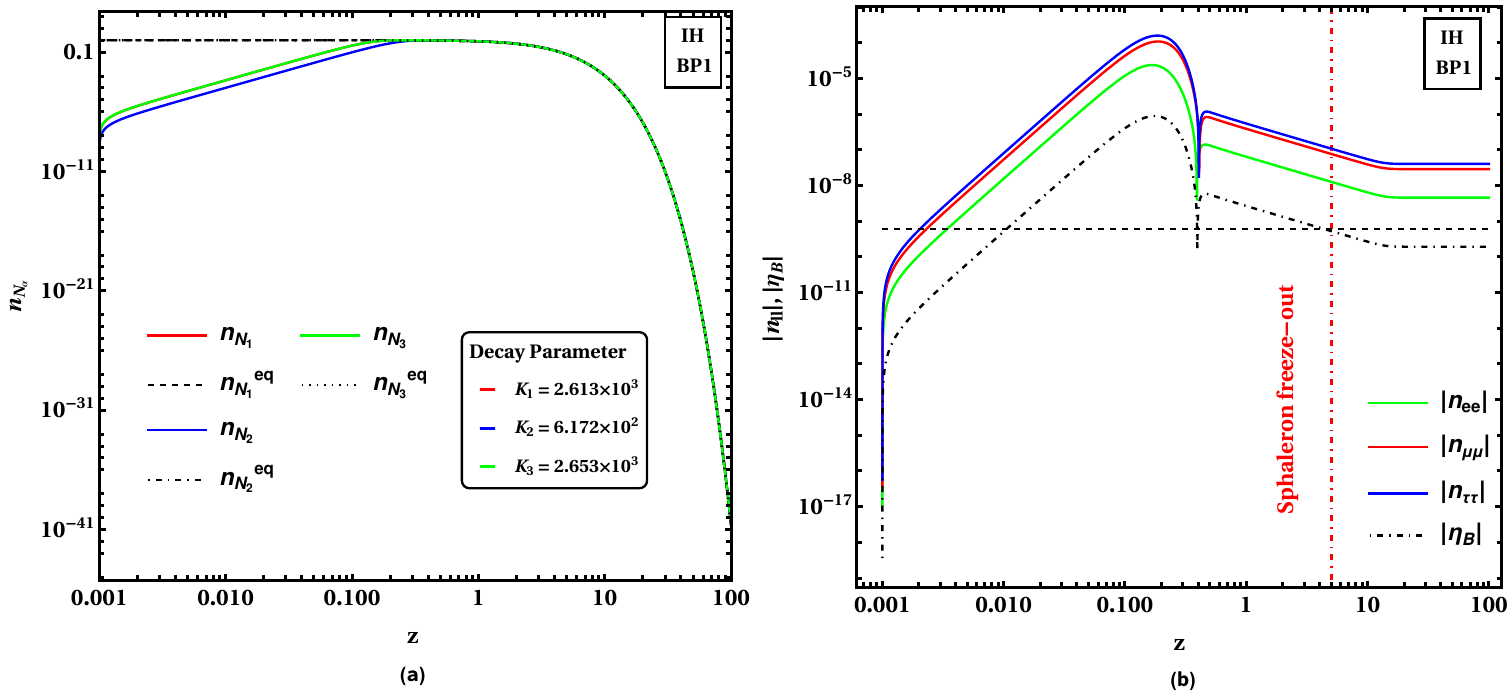}\\
    \includegraphics[scale=0.5]{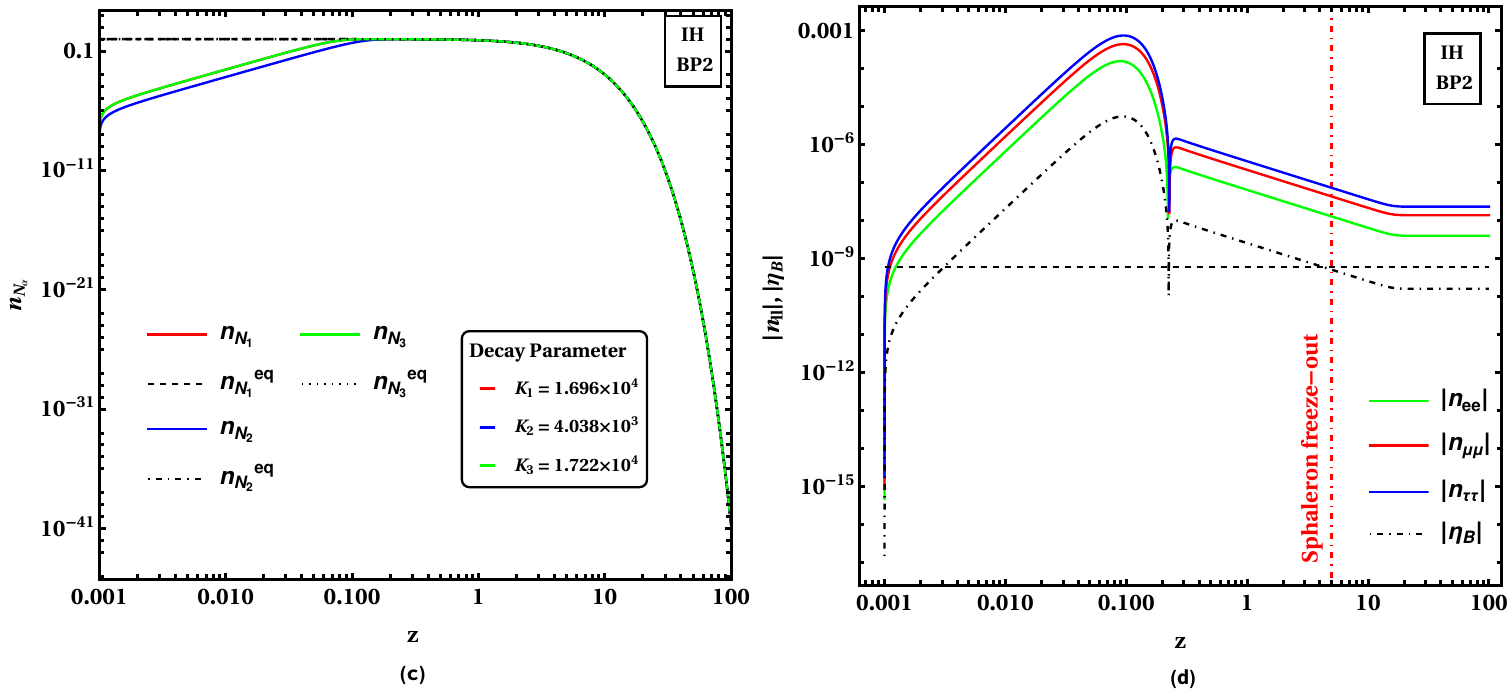}
    \caption{Evolution of the number density of each right-handed neutrino in 
Fig.~\ref{fig:IHB}(a) (\ref{fig:IHB}(c)) and evolution of the number 
density of each lepton flavor together with the baryon asymmetry in 
Fig.~\ref{fig:IHB}(b) (\ref{fig:IHB}(d)) as a function of $z$ for BP1 (BP2) 
in the normal hierarchy (NH) case. The horizontal dashed black line and 
the vertical dot-dashed red line in Fig.~\ref{fig:IHB}(b) and 
Fig.~\ref{fig:IHB}(d) indicate the experimentally observed value of the 
baryon asymmetry and the sphaleron freeze-out, respectively. Zero initial 
abundance is assumed for all particle species.
}
    \label{fig:IHB}
\end{figure}

In addition to the study of  successful baryogenesis  at low energy, in our model, we also aim to investigate the relationship between the required mass degeneracy  needed to obtain the correct baryon asymmetry  and the strong washout regime.
For this purpose, we consider two types of benchmark points BP1 and BP2 for each mass hierarchy (\textit{i.e.,} NH and IH), classified according to the strength of the washout of the generated asymmetry. The BP1 and BP2 crossponds to  low and high  decay parameter strength relative to each other, respectively. The benchmark points, along with the corresponding decay parameters ($K_\alpha$), are presented in Table~\ref{tab:BP}. 
The  results obtained are shown in Fig. \ref{fig:NHB}  (Fig. \ref{fig:IHB}) for NH (IH), where first and second row crossponds to BP1 and BP2, repectively. We consider zero initial abundances for both right-handed neutrinos and leptons. 
In the case of normal hierarchy (NH), BP1 corresponds to a right-handed (RH) 
neutrino mass of approximately $537$~GeV while for BP2 the mass is around 
$592$~GeV. Figure~\ref{fig:NHB}(a) (Fig.~\ref{fig:NHB}(c)) shows the evolution 
of the number densities of each RH neutrino species along with their 
corresponding equilibrium values for BP1 (BP2). The decay 
parameter is inversely related to $z_{\text{eq}}$~\cite{Buchmuller:2004nz}, defined as the value of $z$ at which $n_{N_\alpha}(z_{\text{eq}}) = n_{N_\alpha}^{\text{eq}}(z_{\text{eq}})$.
This behavior is clearly visible in the figures, a larger decay parameter 
corresponds to a smaller value of $z_{\text{eq}}$ for the corresponding RH 
neutrino species.
For $z > z_{\text{eq}}$, the RH neutrino number densities closely follow their 
equilibrium values, indicating that the system lies in the strong washout 
regime. From Fig.~\ref{fig:NHB}(b) (Fig.~\ref{fig:NHB}(d)) for BP1 (BP2), we 
observe that as $n_{N_\alpha}$ in Fig.~\ref{fig:NHB}(a) 
(Fig.~\ref{fig:NHB}(c)) approaches its equilibrium value, the lepton asymmetry 
in different flavors, $|n_{\ell\ell}|$, also approaches its maximum value. 
Subsequently, the lepton asymmetry is washed out and begins to decrease. In the region below the sphaleron freeze-out point $z_{\text{sph}}$, indicated by the red vertical dot-dashed line, the generated lepton asymmetry can be partially converted into the baryon asymmetry of the Universe. The final baryon asymmetry is therefore fixed at $z = z_{\text{sph}}$, beyond which sphaleron processes become ineffective and no further conversion is possible. This final value corresponds to the baryon asymmetry observed today. As can be seen from the Figs. \ref{fig:NHB}(b) and \ref{fig:NHB}(d), where evolution of baryon asymmetry $|\eta_B|$ is shown in black dot-dashed line, we successfully obtain the observed baryon asymmetry, $\eta_B = 6.12 \times 10^{-10}$, for both benchmark points at $z=z_{\text{sph}}$.

As evident from the data in Table~\ref{tab:BP}, the required degeneracy $\delta_{ij}/m_{N_{i}}$ among RH neutrino masses for BP1 is of order $\mathcal{O}(10^{-7} - 10^{-6})$, whereas for BP2 it is $\mathcal{O}(10^{-9})$. This difference arises because BP2 lies deep in the strong washout regime. Consequently, achieving the correct baryon asymmetry requires a larger  total CP asymmetry $\varepsilon_T$ (see  Table~\ref{tab:BP}), which in turn necessitates being closer to the resonant condition. It is also noteworthy that, despite having Yukawa couplings of order $\mathcal{O}(10^{-5} - 10^{-4})$, the correct baryon asymmetry can still be obtained for RH neutrino masses as low as $537$~GeV. We also noted that in order to have observed baryon asymmetry value we need not to go very near the resonance and thus fine tuning is not required.

In the case of inverted hierarchy (IH), BP1 corresponds to  RH neutrino mass 
of $572$~GeV, while BP2 corresponds to a mass of $557$~GeV. 
Figure~\ref{fig:IHB}(a) (Fig.~\ref{fig:IHB}(c)) shows the evolution of RH  neutrino number densities for BP1 (BP2), while Fig.~\ref{fig:IHB}(b) (Fig.~\ref{fig:IHB}(d)) shows the evolution of the generated lepton asymmetry and the final baryon asymmetry for BP1 (BP2). The qualitative behavior in the IH case is similar to that discussed for NH and is therefore not repeated here.
However, it is important to note that  for BP1 in IH case, the degeneracy  $\delta_{ij}/m_{N_i}$ among RH neutrinos required for successful baryogenesis is of the order $\mathcal{O}(10^{-8})$ (see Table \ref{tab:BP}). This is because the total CP asymmetry $\varepsilon_T$ value cannot be very large  due to the small CP violation, as evident from the values of the CP invariants and CP phases discussed in the previous section. In fact, only the Majorana phase $\alpha_{23}$ deviates significantly from trivial values. As a result, one must move closer to the resonance to obtain the correct baryon asymmetry. For BP2 in the IH case, the required degeneracy is of order $\mathcal{O}(10^{-9})$. In this case also we have successfully obtained the baryon asymmetry value, $\eta_B=6.12 \times 10^{-10}$ for both benchmark points at $z=z_{\text{sph}}$. We also noted that, it is very difficult to obtain the experimentally observed baryon asymmetry in IH case if we consider point with higher decay parameter even if we go very close to resonance.

\section{Conclusions}\label{sec6}
In this work, we have successfully implemented the scotogenic model within the framework of non-holomorphic $A_4$ modular symmetry. We have demonstrated that this framework naturally accommodates three nearly degenerate right-handed (RH) neutrinos. This feature arises when the symmetric contribution to the Majorana mass matrix, originating from the $\mathbf{3} \otimes \mathbf{3}$ decomposition of the $A_4$ group representation is treated as a small perturbation to the dominant contribution from the singlet component of the same product.
We have identified regions of parameter space that are consistent with current neutrino oscillation data. In this model, the complex modulus $\tau$ is the only complex parameter and therefore serves as the sole source of CP violation. For both normal (NH) and inverted (IH) mass hierarchies, $\tau$ is restricted to a very narrow region in the complex plane making the model highly predictive and leading to  testable consequences.
In the NH case, we have sizable CP violation. While it is relatively small with only Majorana phase $\alpha_{23}$ taking non-trivial values in IH. 
% This qualitative distinction between the two hierarchies constitutes one of the robust predictions of the model.
Additional hierarchy-dependent predictions arise for the atmospheric mixing angle $\theta_{23}$, the effective Majorana mass $m_{ee}$, and the sum of neutrino masses $\sum_i m_i$.
For NH, the atmospheric mixing angle $\theta_{23}$ is not strongly constrained. The effective Majorana mass $m_{ee}$ lies below the projected sensitivity of upcoming neutrinoless double beta decay ($0\nu\beta\beta$) experiments, while the sum of neutrino masses $\sum_i m_i$ remains safely below current cosmological upper bounds. In contrast, for the IH case, $\theta_{23}$ is predicted to be close to maximal mixing, lying in the lower octant within the range $\theta_{23} \in [44.50^\circ,\,45.04^\circ]$. The predicted values of $m_{ee}$ fall within the sensitivity reach of KamLAND-Zen and future $0\nu\beta\beta$ experiments. Moreover, $\sum_i m_i$ lies close to the present upper limit from Planck observations, making it testable by forthcoming cosmological measurements.
Furthermore, we have studied tri-resonant leptogenesis within this model and successfully generated the observed baryon asymmetry of the Universe at a low energy scale, with RH neutrino masses as low as $537~\mathrm{GeV}$. This opens the possibility of probing the model in current and near-future experiments~\cite{Antusch:2017pkq,Anamiati:2016uxp}. We find that baryogenesis is viable for both neutrino mass hierarchies. However, in the NH case, a mass degeneracy of order $\mathcal{O}(10^{-7}\text{--}10^{-6})$ among the RH neutrinos is sufficient, whereas in the IH case a stronger degeneracy of order $\mathcal{O}(10^{-8})$ is required. Notably, in the NH scenario, even with a decay parameter as large as $\mathcal{O}(10^{5})$, the correct baryon asymmetry can be achieved due to the tri-resonant enhancement of the CP asymmetry and the inclusion of flavor effects.
Thus, the framework studied here is highly predictive and offers multiple avenues for experimental verification. The precise measurements of the atmospheric mixing angle $\theta_{23}$, improved determination of CP-violating phases, next-generation $0\nu\beta\beta$ decay searches, and refined cosmological bounds on the sum of neutrino masses have strong potential to probe and possibly exclude large regions of the model’s parameter space.

% \begin{figure}[ht]
%     \centering
%     \includegraphics[width=0.45\textwidth]{nf_asy_vs_z.pdf}
%     \includegraphics[width=0.45\textwidth]{B_asy_vs_z.pdf}

%     % \includegraphics[width=0.45\textwidth]{M32_vs_CP.pdf}
%     % \includegraphics[width=0.45\textwidth]{fig4.pdf}

%     \caption{Evolution of number density of each flavor (left panel) and baryon asymmetry  (right panel) as a function of z. We have considered zero initial abundance for all the particles. }
% \end{figure}

% \begin{figure}[h]
%     \centering
%     \begin{tikzpicture}
%         \begin{feynman}
%             % Define vertices
%             \vertex (a) {\(\nu_L\)};
%             \vertex [right=2.5cm of a] (b);
%             \vertex [below right=2.5cm and 1.5cm of b] (c);
%             \vertex [above right=2.5cm and 1.5cm of b] (d);
%             \vertex [right=2.5cm of c] (e) {\(\nu_L\)};
            
%             % Draw external fermion lines
%             \diagram* {
%                 (a) -- [fermion] (b),
%                 (b) -- [fermion, edge label=\(N_R\)] (c),
%                 (c) -- [scalar, edge label'=\(\eta\)] (d),
%                 (d) -- [fermion, edge label'=\(N_R\)] (b),
%                 (c) -- [fermion] (e),
%             };
            
%             % Add labels for the scalar and fermion loop
%             \node at ($(b)!0.5!(c)$) [above left] {\(N_R\)};
%             \node at ($(c)!0.5!(d)$) [below right] {\(\eta\)};
%         \end{feynman}
%     \end{tikzpicture}
%     \caption{One-loop neutrino mass generation in the type-I Scotogenic model.}
% \end{figure}

\section*{Acknowledgments}
\vspace{.3cm}
\noindent Tapender acknowledges the financial support provided by Central University of Himachal Pradesh in the form of freeship. The authors, also, acknowledge Department of Physics and Astronomical Science for providing necessary facility to carry out this work.

\bibliographystyle{unsrt}
\bibliography{ref}
\end{document}